\documentclass[%
 aip,
rsi,%
 amsmath,amssymb,
 reprint,%
]{revtex4-1}
\usepackage{graphicx}
\usepackage{amsfonts} 
\usepackage{xcolor}
\usepackage{multirow}
\usepackage{tikz}
\usetikzlibrary{calc}
\tikzset{fontscale/.style = {font=small}}

\begin{document}
\title[A Review of Compact Interferometers]{A Review of Compact Interferometers}

\author{Jennifer Watchi}
	\email{jwatchi@ulb.ac.be}
	\affiliation{Precision Mechatronics Laboratory, BEAMS Department, ULB, BE}

\author{Sam Cooper}%
\affiliation{School of Physics and Astronomy and Institute of Gravitational Wave\\ Astronomy, University of Birmingham, Birmingham B15 2TT, UK}

\author{Binlei Ding}
	\affiliation{Precision Mechatronics Laboratory, BEAMS Department, ULB, BE}

\author{Conor M. Mow-Lowry}%
\affiliation{School of Physics and Astronomy and Institute of Gravitational Wave\\ Astronomy, University of Birmingham, Birmingham B15 2TT, UK}

\author{Christophe Collette}
	\affiliation{Precision Mechatronics Laboratory, BEAMS Department, ULB, BE}
	\affiliation{Department of Aerospace and Mechanical Engineering, ULg, BE}

\date{\today}

\begin{abstract}
Compact interferometers, called phasemeters, make it possible to operate over a large range while ensuring a high resolution. Such performance is required for the stabilization of large instruments dedicated to experimental physics such as gravitational wave detectors. This paper aims at presenting the working principle of the different types of phasemeters developed in the literature. These devices can be classified into two categories: homodyne and heterodyne interferometers. Improvement of resolution and accuracy has been studied for both devices. Resolution is related to the noise sources that are added to the signal. Accuracy corresponds to distortion of the phase measured with respect to the real phase, called non-linearity. The solutions proposed to improve the device resolution and accuracy are discussed based on a comparison of the reached resolutions and of the residual non-linearities.
\end{abstract}



\maketitle 

\section{Introduction}
Relative motion between two points can be measured by a number of transducers, converting the variation of a physical quantity into some useful voltage. Some examples of commonly used sensors are capacitive sensors, linear variable differential transformers (LVDT) and eddy current sensors. For each application, the adequate choice depends on many criteria, including resolution, dynamic range, space available, price, and compatibility with operating environment. While based on very different working principles, all of theses sensors are fundamentally limited by a trade-off between resolution and dynamic range. In other words, none of them can process both small and large quantities. Moreover, even the most sensitive of these techniques have limited resolution, and are not reliable in operating environments with stray magnetic fields.

These two aforementioned limitations prevent them from being used in many applications like high precision machine tools or production chains.

Interferometers are an excellent alternative due to their high sensitivity, non-contact measurement, and immunity to magnetic couplings. Conventional interferometers have a small working range, but when the optical phase is measured in two quadratures, the output can be unwrapped creating a large working range optical-phasemeter. 

Compact optical-phasemeters are of increasing interest to physics and precision engineering communities. In this paper we review a range of devices that can be called `compact', which implies that the interferometer is an enabling tool and that either the complete system or an optical head can be deployed onto an apparatus. While not all reviewed works clearly specify the size and form of the interferometer, we have attempted to apply these two criteria to determine their relevance. For convenience, we often refer to the complete signal chain from the interferometer to the unwrapped phase readout simply as a phasemeter.

Many prototypes of compact interferometers have been developed for two principal types of applications. The first application is as a simple position sensor. Such sensors have been used in gravitational wave detectors for local damping~\cite{Aston2011} or on the ISI~\cite{strain2012damping,voyager}.  The second application is in the development of high-resolution inertial sensors, where one mirror is fixed on an inertial mass~\cite{Zumberge04}. These sensors are useful for the stabilization of gravitational wave detectors \cite{MatichardLantz2015, Venkateswara15}, gravimeters \cite{Pena13,Harms16Notes, Zhou15, Zhou12, McGuirk01} or particle accelerators  \cite{Collette12}. 

The objective of this paper is to provide a comparison of compact interferometers in terms of resolution, dynamic range, and linearity. The focus is on devices with a working range of more than one fringe. The paper starts with a brief section explaining the working principle and limitations of conventional two-beam and resonant interferometers. It is followed by Sections \ref{sec:homodyne} and \ref{sec:heterodyne} dedicated to homodyne- and heterodyne-phasemeters. For each of them, the working principle is presented and several examples from relevant literature are described.

Sections \ref{sec:nonlinearities} and \ref{sec:noisesources} discuss problems that are common to all types of  phasemeters, and counter measures that mitigate these problems. The first is the limited accuracy due to the non-linearities in the phase measurement. The second is a short review of the main noise sources in interferometers. 
The paper concludes with historical trends, and a discussion on the dimensions of compact interferometers.

\section{Small range interferometers}
\label{sec:smallRange}

The focus of this paper is on large-range interferometers, capable of tracking the position of a target mirror with resolution much smaller than a wavelength over a working range of (much) more than a wavelength. In this section the key interferometry concepts and nomenclature are introduced with examples for standard small-range (sub-wavelength) interferometers. We consider two-beam interferometers, such as Michelson, Mach-Zender, and Sagnac interferometers, separately from resonant (or multi-bounce) interferometers. The function of actuators to increase the working range of devices is briefly introduced.

The standard nomenclature for analysing laser-interferometers is a form of short-hand that simplifies the electric field into a single-sided, complex function that is integrated over the transverse profile and re-normalised such that the power, $P$, of a beam is the mod-square of the field, $E$. The complex form is especially useful since the field can be represented by phasors, and interference as the vector sum of phasors. Mirrors (and beam splitters) can then be treated as having a field reflectivity, $r$, that is the square root of the power reflectivity, $R$, and a field transmission of $t = \sqrt{T} = \sqrt{1 - R}$. We use the convention that a phase shift of $i$ is gathered after transmission through an interface. An excellent introduction to interferometry, including nomenclature, can be found in Ref.~\cite{Bond2017}. 

The transverse profile of the electric field is not considered in this section, an approximation that is valid when both the paraxial approximation holds and when all interfering beams have significant spatial overlap efficiency,  greater than $\sim$10\,\%. Details on transverse modes, and their interations with resonators, can be found in, for example, Refs. \cite{Siegman, Saleh, gossler2003}. 

\subsection{Two-beam interferometers}
\label{sec:twobeam}
\begin{figure}[ht]
	\centering
	\includegraphics[trim = {0cm 0.5cm 0cm 1cm},width=0.5\textwidth]{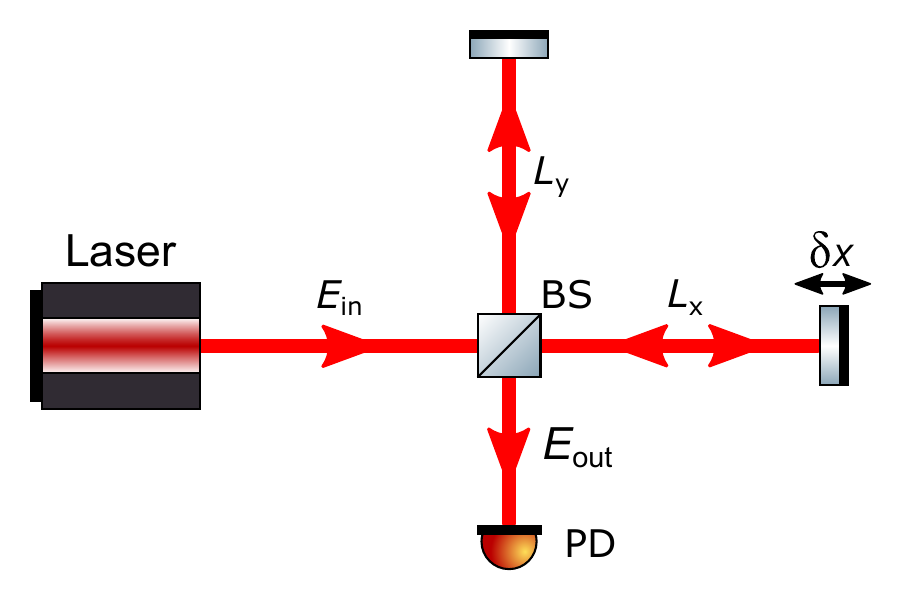}
	\caption{A simple Michelson interferometer with input and output fields $E_{\rm in}$ and $E_{\rm out}$, with a (non-polarising) beamsplitter (BS) of power-reflectivity R, and arms of length $L_{\rm x}$ and $L_{\rm y}$. The power measured on the photodiode (PD) is dependent on the phase shift acquired in the arms.}
	\label{fig:simpleMich}
\end{figure}

For the Michelson interferometer shown in Fig. \ref{fig:simpleMich} the output field is
\begin{eqnarray}
E_{\rm out} & = i r t E_{\rm in} ( e^{i \phi_x} + e^{i \phi_x} ),
\end{eqnarray}
where $\phi_{x, y}$ is the round-trip phase acquired in the respective arm. It is useful to express this in terms of the sum ($\phi_{\rm s}$) and difference ($\phi_{\rm d}$) of the phases, such that
\begin{eqnarray}
\phi_x  = \frac{\phi_{\rm s} + \phi_{\rm d}}{2}, \qquad \phi_y = \frac{\phi_{\rm s} - \phi_{\rm d}}{2}.
\end{eqnarray}
Assuming the beam splitter is lossless and has $R = T = 0.5$, the output power, $P_{\rm out} = |E_{\rm out}|^2$, as a function of the input power, $P_{\rm in}$, is
\begin{eqnarray}
P_{\rm out} & =  \dfrac{P_{\rm in}}{2} (1 + \cos(\phi_{\rm d})),
\label{eq:mich}
\end{eqnarray}
This dependence of power on differential phase applies generally to two-beam interferometers, including Sagnac and Mach-Zender devices, although the fringe visibility may be affected by the beam splitter ratio. However in some configurations, such as the Sagnac interferometer, the coupling of displacement (or laser frequency) to differential phase is substantially different, and the analysis below is limited to the Michelson interferometer.

With Eq. \ref{eq:mich} we see that the output power is independent of the sum (or common) arm length. For a monochromatic light source with wavelength $\lambda$, and wavenumber $k = 2 \pi / \lambda$, the optical phase difference is simply
\begin{eqnarray}
\phi_{\rm d} = 2 k \Delta L,
\end{eqnarray}
proportional to the arm length difference, $\Delta L = L_x - L_y$. Since the output power is sinusoidal, at the turning points the sensitivity to length goes to zero, and the direction of motion becomes ambiguous. For these reasons, normal two-beam interferometers that measure the output power have a small operating range of less than half a wavelength. However, later sections will show that it is possible to extract the optical phase by using a combination of additional optical components and signal processing to produce a {\it phasemeter} instead of an interference-meter.

Since even narrow linewidth lasers are not monochromatic, and frequency fluctuations are often a significant source of noise in many precision interferometry experiments, it is useful to determine how frequency fluctuations couple to the optical phase. We can do this by separating the wavenumber into an average component, $k_0$, and a time-fluctuating  component, $\delta k$. The length difference is similarly divided into constant, $L_0$, and fluctuating $\delta \! L$ components. In both cases, the time-fluctuating component is assumed to be much, much smaller than the constant value. Combining these terms, the differential phase is now:
\begin{eqnarray}
\phi_d & = & 2 (k_0 + \delta k)(L_0 + \delta \! L) \\
 & \approx & 2(k_0 L_0 + k_0 \delta \! L + \delta k L_0),
\end{eqnarray}
where the three terms in the second line are: the static offset of the interferometer (sometimes called the `operating point' or `tuning'); the length signal; and the frequency fluctuations ($\delta k = 2 \pi \delta \! f / c$ for frequency fluctuations $\delta \! f$) coupling to differential phase. For a commercial, free-running Nd:YAG 1064\,nm laser, the frequency noise is approximately $10^4$\,Hz/$\sqrt{\rm Hz}$ at 1\,Hz, with a characteristic $1/f$ slope \cite{fritschel89}.


\subsection{Optical resonators}

\begin{figure}[ht]
	\centering
	\includegraphics[trim = {0cm 0.5cm 0cm 0.7cm},width=0.5\textwidth]{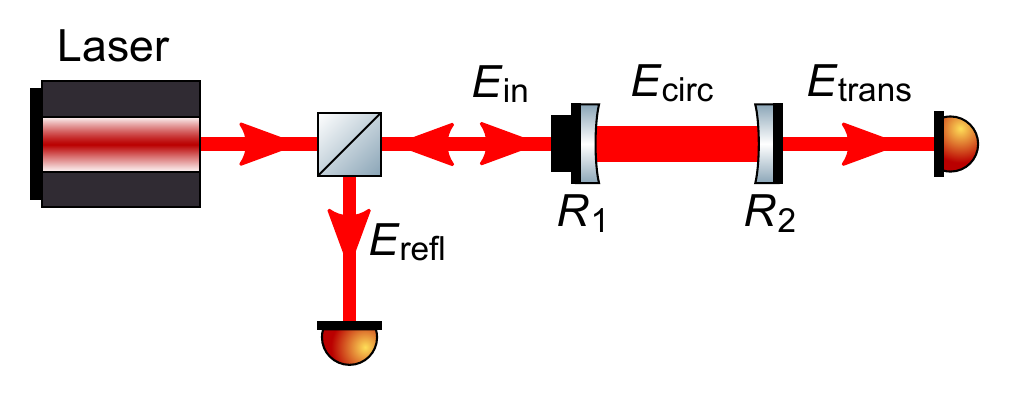}
	\caption{A 2-mirror cavity forming an optical resonator. The labels indicate the input, circulating, transmitted, and reflected optical fields. The two mirrors have power reflectivity $R_1$ and $R_2$.}
	\label{fig:cavity}
\end{figure}

In its standard form, an optical resonator consists of two mirrors, as shown in Fig.\,\ref{fig:cavity}. It is conceptually simple to analyse a resonator as a multiple `bounce' system \cite{Hecht}. In this picture, light is transmitted through the mirror, circulates around the cavity, and interferes with the time-delayed incoming light. The field is vector-summed until a steady-state solution is reached. Resonator quality can be quantified by the effective number of bounces required to reach steady-state, but it is most typically defined by the finesse, $\mathfrak{F}$, which is the ratio of the linewidth (or full-width at half-maximum height, FWHM) of the resonator to the free spectral range (FSR) \cite{Bond2017}
\begin{eqnarray}
\mathfrak{F} = \dfrac{{\rm FSR}}{{\rm FWHM}} \approx \dfrac{\pi \sqrt{r_1 r_2}}{1 - r_1 r_2},
\end{eqnarray}
where the approximation is valid for two-mirror resonators with high reflectivity mirrors, $T_1, T_2 << 1$. 

If the phasors for all the packets inside the cavity add coherently, the circulating field will increase until the power lost in each round trip is equal to the input power.  This condition defines resonance - when the circulating field is at its maximum for a given input field.

There are two requirements for a cavity to be on resonance: the field must self-reproduce spatially (the transverse mode condition), and the circulating field must interfere constructively with the input field (the longitudinal mode condition). 

The response time (the inverse of the linewidth) of small resonators is typically very fast ($10^{-6}$ to $10^{-10}$\,s) compared with the time-scales in most sensing applications (typically longer than $10^{-5}$\,s), and as such resonator can be assumed to be in a steady-state. The cavity fields can then be determined using a set of self-consistent equations \cite{Siegman}. These equations are derived in an intuitive way from the fields shown in Fig. \ref{fig:cavity}.  Note that the input and reflected fields are defined immediately to the left of $R_1$, propagating right and left respectively. The circulating field is defined immediately to the right of $R_1$, propagating to the right. For a lossless system with round-trip phase $\phi$, the fields are:
\begin{eqnarray}
    E_{\rm refl} & = & r_{1} E_{\rm in} + i r_{2} t_{1} E_{\rm circ} e^{i\phi} \\
    E_{\rm circ} & = & i t_{1} E_{\rm in}  + r_{2} r_{1} E_{\rm circ} e^{i\phi}  \\
    E_{\rm trans} & = & i t_{2} E_{\rm circ} e^{i\phi/2},
    \label{eq:cavrelations}
\end{eqnarray}
where $r_n = \sqrt{R_n}$, $t_n = \sqrt{1 - R_n}$, $n = 1,2$. Solving in terms of the input field gives:
\begin{eqnarray}
    E_{\rm refl} & = & E_{\rm in} \dfrac{r_{1} - r_{2} e^{i\phi}}{1 - r_{1} r_{2} e^{i\phi}} \label{eq:cavref} \\
    E_{\rm circ} & = & E_{\rm in} \dfrac{i t_{1}}{1 - r_{1} r_{2} e^{i\phi}} \label{eq:cavcirc} \\
    E_{\rm trans} & = & E_{\rm in} \dfrac{-t_{1} t_{2} e^{i\phi/2}}{1 - r_{1} r_{2} e^{i\phi}}. \label{eq:cavtrans}
\end{eqnarray}

\begin{figure}[ht]
	\centering
	\includegraphics[width=0.5\textwidth]{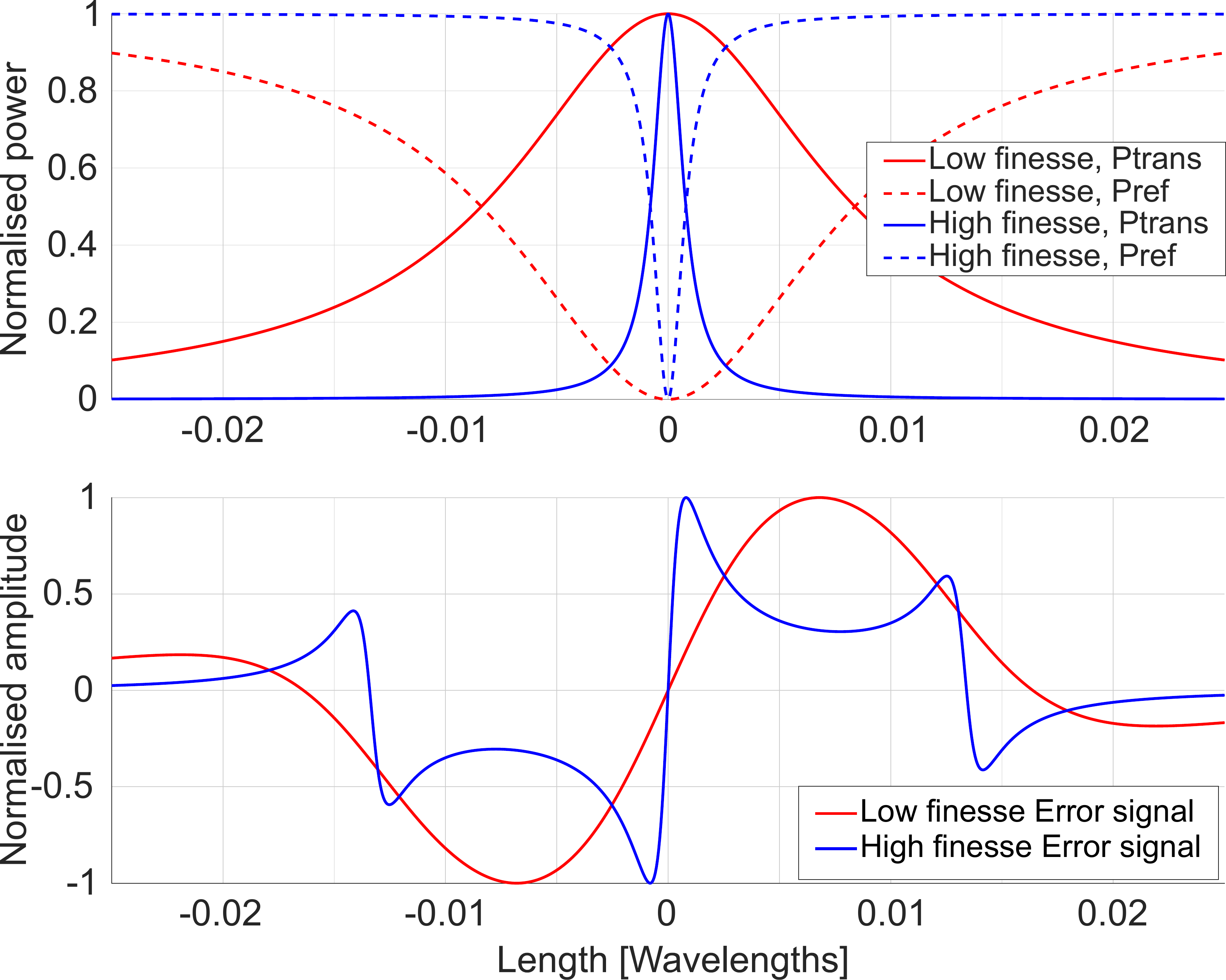}
	\caption{The reflected and transmitted power for lossless, impedance-matched, low- and high-finesse cavities (30 and 300 respectively) along with the error signal produced with the Pound-Drever-Hall technique.}
	\label{fig:cavPlot}
\end{figure}

The upper plot in Fig. \ref{fig:cavPlot} shows the transmitted and reflected power for low- and high-finesse cavities (30 and 300 respectively). As with the two-beam interferometer, if one observes only the power, there is a loss of signal and ambiguity at the turning point. It is possible to operate offset from the centre of resonance (sometimes called side-fringe or mid-fringe readout) such that the power has a well-defined gradient \cite{Barger1973}.

The most common technique used in precision cavity readout is the Pound-Drever-Hall (PDH) technique \cite{Drever1983, Black2001}, where the laser beam is phase-modulated at radio frequencies producing an `error' signal that is dependent on the optical phase when the laser is close to resonance. Typical PDH error signals, calculated using equations in section IV of Ref.~\cite{Black2001}, are shown in the lower plot in Fig.~\ref{fig:cavPlot}, although the low-finesse case is not strictly PDH since the modulation frequency is comparable to the cavity linewidth.

Equations \ref{eq:cavref} and \ref{eq:cavtrans} relate the outgoing fields to the optical phase of the resonator, and this can be converted into power and length in a similar fashion to a two-beam interferometer. It is also  common to measure the laser frequency (or wavelength) rather than the phase~\cite{gerberding15,zhu15fp,zhu15subnanometer}, although relative length fluctuations can be simply equated to the relative frequency and relative wavelength fluctuations by:
\begin{eqnarray}
	\dfrac{\delta \! L}{L} = \dfrac{\delta \! f}{f} = \dfrac{\delta \lambda}{\lambda}, \label{eq:cavStrain}
\end{eqnarray}
assuming that $\delta L \ll L$. Long cavities are therefore better frequency discriminators while short cavities are less effected by frequency fluctuations.

Optical resonators are commonly employed in sensing applications where the multiple bounces from the mirrors amplifies the optical phase-shift. They can also be used to simplify the optical construction by reducing the number of elements. Sensing resonators can be both free-space~\cite{Bhatia96}, where the displacement of one optic changes the path-length, or solid-state (such as fibre-resonators), where the dominant effect is typically stress-induced refractive index changes. The increase in sensitivity compared with two-beam interferometers comes at the expense of the working range, which is smaller by a factor of approximately $\mathfrak{F}$.


\subsection{Actuators to increase working range}

In most practical applications, small-range interferometric sensors are operated using closed-loop feedback to hold them within their working range. There are several possible mechanisms. The length of the reference arm can be altered, for example with a piezo-electric transducer\,\cite{Gray1999}. The readout of the target mirror position is then encoded in the actuation voltage to the piezo, and the dynamic range is limited by the driving electronics, which can be up to 9 orders of magnitude.

The laser frequency (or wavelength) can also be controlled, and the phase change of the interferometer is extracted in the frequency actuation. A recent example uses a laser with a traceable wavelength calibration to link acceleration measurements with existing standards\,\cite{Guzman14}. 

Alternatively it is possible to act on the target mirror, creating a complete device that is operationally similar to force-feedback seismometers \cite{Wielandt82,Xie10}. In all these cases, the dynamic range and linearity of complete system is limited by the actuation mechanism, and any intrinsic noise in the actuator must be considered. In contrast, phasemeters use fringe-counting in signal processing, which in principle has a dynamic range only limited by numerical precision and the tracking speed of the fringe-counter. This allows phasemeters to use the full dynamic range of the readout electronics (typically limited by the ADC) for each fringe.

The use of actuators is important for extending the range of interferometric readout, but the limitations, linearity, and range of closed-loop actuator-readout is beyond the scope of this review. The following sections focus on optical readout based on phasemeters that have inherently large working range.

\section{Homodyne Phasemeters}
\label{sec:homodyne}


To increase the working range of a two-beam interferometer, the phase must be unambiguously extracted over more than one cycle, which is not possible by using Eq. \ref{eq:mich}. To increase the interferometer dynamic range the general idea consists of creating two signals in quadrature, $P_1$ and $P_2$, given by 
\begin{eqnarray}
	P_1 & = & P_{0}(1 + \cos (\phi_{\rm d})), \\
	P_2 & = & P_{0}(1 + \sin (\phi_{\rm d})),
	\label{eq:iqSimple}
\end{eqnarray}
where $P_0$ is determined by the optical power and the fraction of it that reaches the sensors. Then, an arbitrarily large phase can be calculated using
\begin{eqnarray}	
	\phi_{\rm d} & = & {\rm atan2}((P_1 - P_0), \, (P_2 - P_0)).
	\label{eq:phaseReadout}
\end{eqnarray}

Since the unwrapping occurs in signal processing, the fringe-counting is noiseless as long as the direction of the wrapping is known. The atan2 function provides the unwrapped phase assuming that it is evaluated on a circle.  For the rest of the section, we will consider the ideal case that corresponds to two perfect quadrature signals. Phase shift issues and any other causes of circle distortion are discussed in the section~\ref{sec:nonlinearities}.\\

In this section, different methods to generate quadrature signals are presented. The quadrature signals can be carried by the two polarizations states of the beam of by two transverse modes of the intensity beam profile.
The advantages and drawbacks of these methods are mainly related to the resolution of the interferometer, which is the smallest physical quantity that a sensor can measure \cite{Collette12}. Here, the smallest physical quantity is the noise of the measurand. The sources of noises will be introduced in section \ref{sec:noisesources}.

\subsection{Quadrature signals carried by the polarization states: additional wave plates}
Two quadrature signals can be generated by imposing a phase shift of $\pi/2$ between the two polarization states of a beam thanks to a waveplate. The phase shift of $\pi/2$ can be obtained either by passing once through a $\lambda /4$ waveplate or twice through a $\lambda/8$ waveplate. The implementation of these two options to obtain quadrature signals is detailed below.

\subsubsection{$\lambda/8$ wave plate}
A $\lambda/8$ wave plate is placed in one of the interferometer's arms to provide a differential (round-trip) phase shift of $\pi/2$ between two linear polarizations~\cite{Downs79,Zumberge04,Otero09}. In fact, this creates two co-located Michelson interferometers, one in each polarization, that measure the target mirror. The outputs of these interferometers are then separated by using a polarizing beam splitter.

\begin{figure}[ht!]
    \centering
    \includegraphics[trim = {1cm 0.3cm 1cm 0.5cm},width=0.4\textwidth]{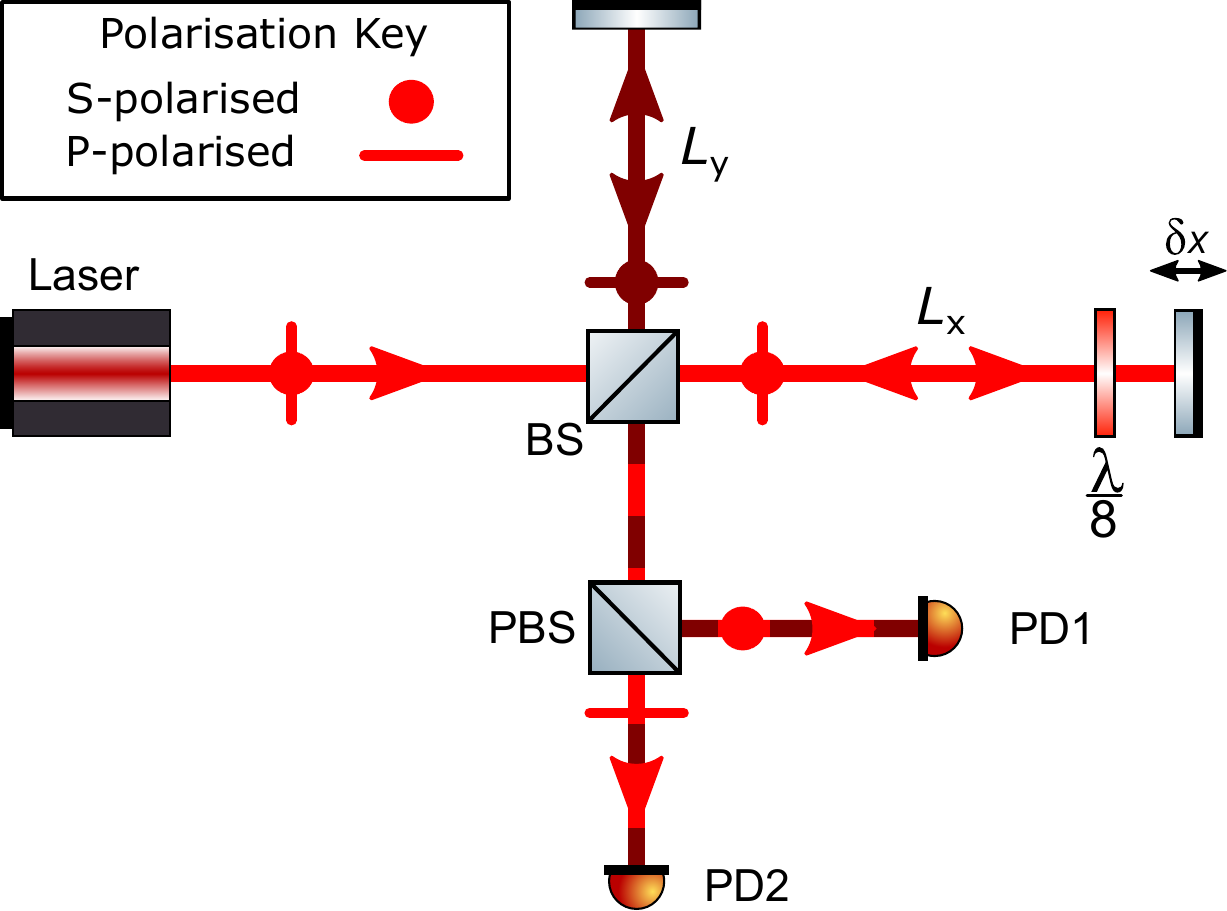}
    \caption{A homodyne phasemeter. The $\lambda/8$ wave plate in the $x$-arm has its fast axis aligned with the s- (or p-) polarisation, effectively creating two co-incident Michelson interferometers with $\pi/2$ different arm lengths. The polarising beamsplitter (PBS) splits the two outputs onto the photodiodes PD1 and PD2.}
    \label{fig:lambda8waveplate}
\end{figure}

A schematic representation is shown in Fig.~\ref{fig:lambda8waveplate} where the dot on the beam indicates the s-polarized axis and the perpendicular line, the p-polarized axis. The beam is split by a non-polarizing beam splitter and then one polarization is delayed in the x-arm. After recombination at the beam splitter, the two polarizations are measured independently at the photodiodes 1 and 2. 

An interferometer of this kind has been mounted in a seismometer~\cite{Zumberge04,Otero09}. It has a resolution of around 1~pm/$\sqrt{\rm Hz}$ at 1~Hz. Several modifications of the optical path have been introduced to reduce noises and hence improve the interferometer resolution. 
These structure modifications are discussed hereafter. 

\paragraph{Extra photodiodes to delete the DC component}
\label{sec:HomodyneExtraphotodiodes}
Fringe counting becomes complicated when the sinusoidal signals are not oscillating around zero. Several methods have been implemented to compensate this shift but they are difficult to apply for noise with a very low frequency~\cite{Downs79}. In Ref.~\cite{Downs79}, three polarizations states are measured by using two polarizing beam splitters instead of one: two out of phase polarizations and one orthogonal polarization are measured. 
If we don't consider a gain mismatch between sensors, 
 the three signals can be written as
\begin{eqnarray}
P_{PD1} &=& P_0(1 +  \sin(\phi_{\rm d})) \\
P_{PD2} &=&  P_0 (1 + \cos(\phi_{\rm d})\\
P_{PD3} &=&  P_0 (1 - \cos(\phi_{\rm d})
\end{eqnarray}
Thanks to a correct subtraction of the sine signal to the two others, the two resulting signals are in quadrature and the DC component is removed:
\begin{eqnarray}
P_1 &=& P_{PD1}-P_{PD2} = \sqrt{2}P_0 \sin(\phi_{\rm d}-\frac{\pi}{4}) \label{eq:P1_3PD}\\
P_2 &=& P_{PD1}-P_{PD3} = \sqrt{2}P_0 \sin(\phi_{\rm d}+\frac{\pi}{4}) \label{eq:P2_3PD}
\end{eqnarray}
As the phase is obtained from the atan2 of the ratio between these two signals, the results become insensitive to the input power fluctuations. 
Consequently, the resolution is not deteriorated even when the laser intensity drops down to 10~\%~\cite{Downs79}.

The use of additional photodiodes also has certain advantages for the reduction of non-linearities which is discussed in Section~\ref{sec:AdditionalSensors}.

\begin{figure}[ht!]
    \centering
	\includegraphics[width=1\linewidth]{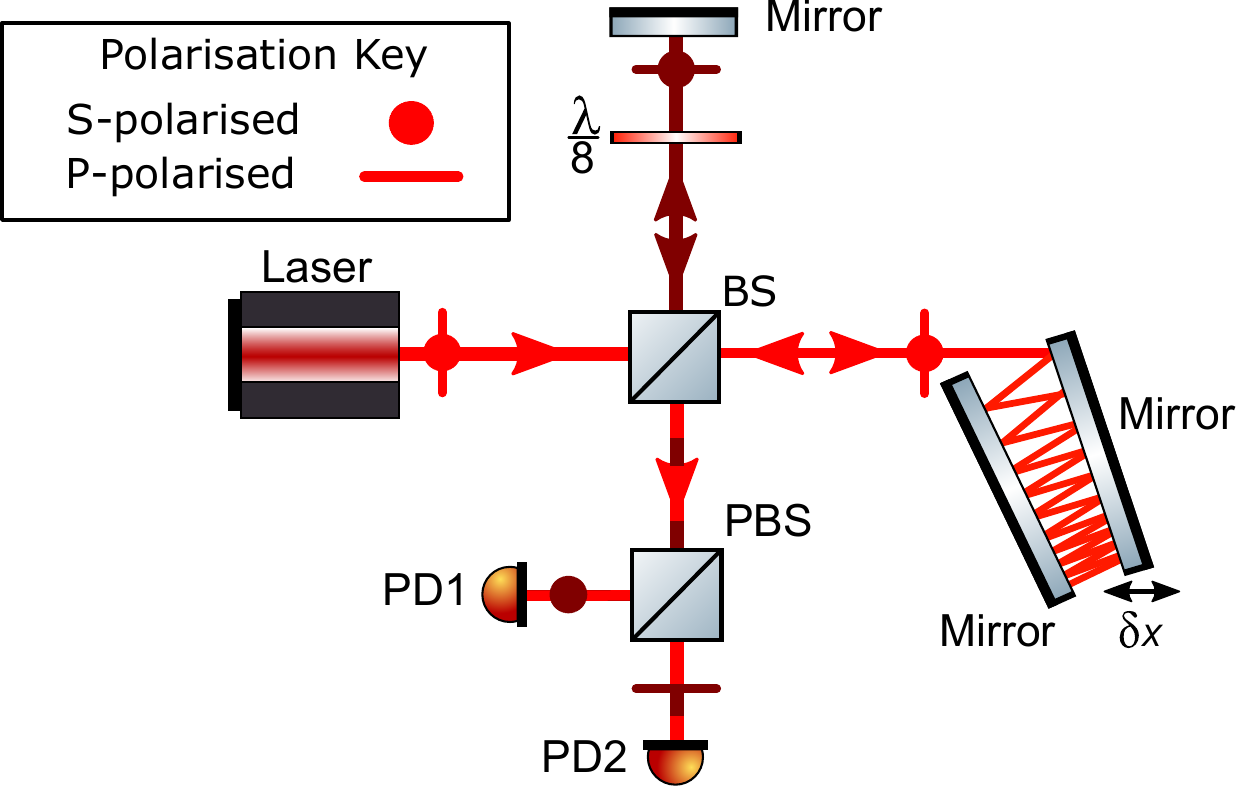}
	\caption{A homodyne phasemeter using a $\lambda/8$ wave plate and multiple reflections on the target mirror to enhance sensivity. Adapted from the experimental setup figure in Ref.~\cite{Pisani09}.}
    \label{fig:Multipass}
\end{figure}

\paragraph{Multiple-reflections in the measurement arm}
\label{sec:MultipassHomodyne}
One way to improve the resolution is to increase the number of reflections on the target mirror by slightly tilting the mirror and placing a fixed mirror in front of it, see Fig. \ref{fig:Multipass}. If the measurement mirror moves along its normal axis, represented by $\delta x$ on the figure, the phase change occurs at each reflection \cite{Pisani09}. Consequently, the phase measured is proportional to $G \delta x$, where $G$ corresponds to the number of reflections on the moving mirror, see Fig. \ref{fig:Multipass}.

Consequently, the smallest phase increment measurable is proportional to $\delta x /G$. 
It means that the resolution is improved of a factor $G$ in comparison with a single-bounce interferometer. In Ref.~\cite{Pisani09}, this assumption has been verified experimentally: a comparison between a simple Michelson interferometer and a 60 reflections version has been presented. Around 2~Hz, the new configuration resolution is 20 times better than the classical version. At high frequencies, an improvement of the resolution by a factor 60 is reached. The resolution can still be improved because while using multiple-reflections in one arm, the phase noise related to unequal optical arm length increases, which is introduced in section \ref{Frequencynoise}. 

To increase the resolution, the number of reflections must be as large as possible. However, the beam should not overshoot the size of the mirror. An optimum number of reflections can be adjusted as explained in Ref.~\cite{zhou01}.
In addition, the number of reflections cannot be too large to avoid being beyond the laser coherence. In order to maintain the coherence between the two paths, a Michelson interferometer with two multiple-reflections arms has been studied~\cite{Joenathan16}. The two mirrors are rotated with the same angle as they are coupled thanks to a gear mechanism. Because the two beams are reflected the same number of times, the intensity losses due to the multiple bounces are also identical. In comparison, in the system with a single multiple-reflections arm, the intensity loss must be estimated because it reduces the fringe visibility~\cite{Joenathan16}.

Finally, for large number of reflections, the environment can induce phase jumps. Therefore, a compromise must be found between an increase in resolution and a loss of coherence. All these aspects and their impact on the delay are discussed in the multiple-reflection interferometers papers~\cite{Pisani09,Joenathan16}.

\paragraph{Other configurations}
Some additional modifications can be found in the literature. Their impact on the resolution is not clear or has not been verified experimentally. In Ref.~\cite{Downs79}, it is suggested that a lens can be used to reduce the beam motion across the active area of the photodiode. Moreover, in Ref.~\cite{reibold1981laser,daniel2005advanced}, the polarizing beam splitter used to separate the two polarizations is replaced by a Wollaston prism.  With this prism, the two polarization states are emitted in the same plane but their direction varies with a defined angle.

\subsubsection{$\lambda /4$ wave plate}
\label{HomodyneSectionB2}
\begin{figure}[ht!]
	\centering
	\includegraphics[trim = {0cm 0cm 0cm 0cm},width=1\linewidth]{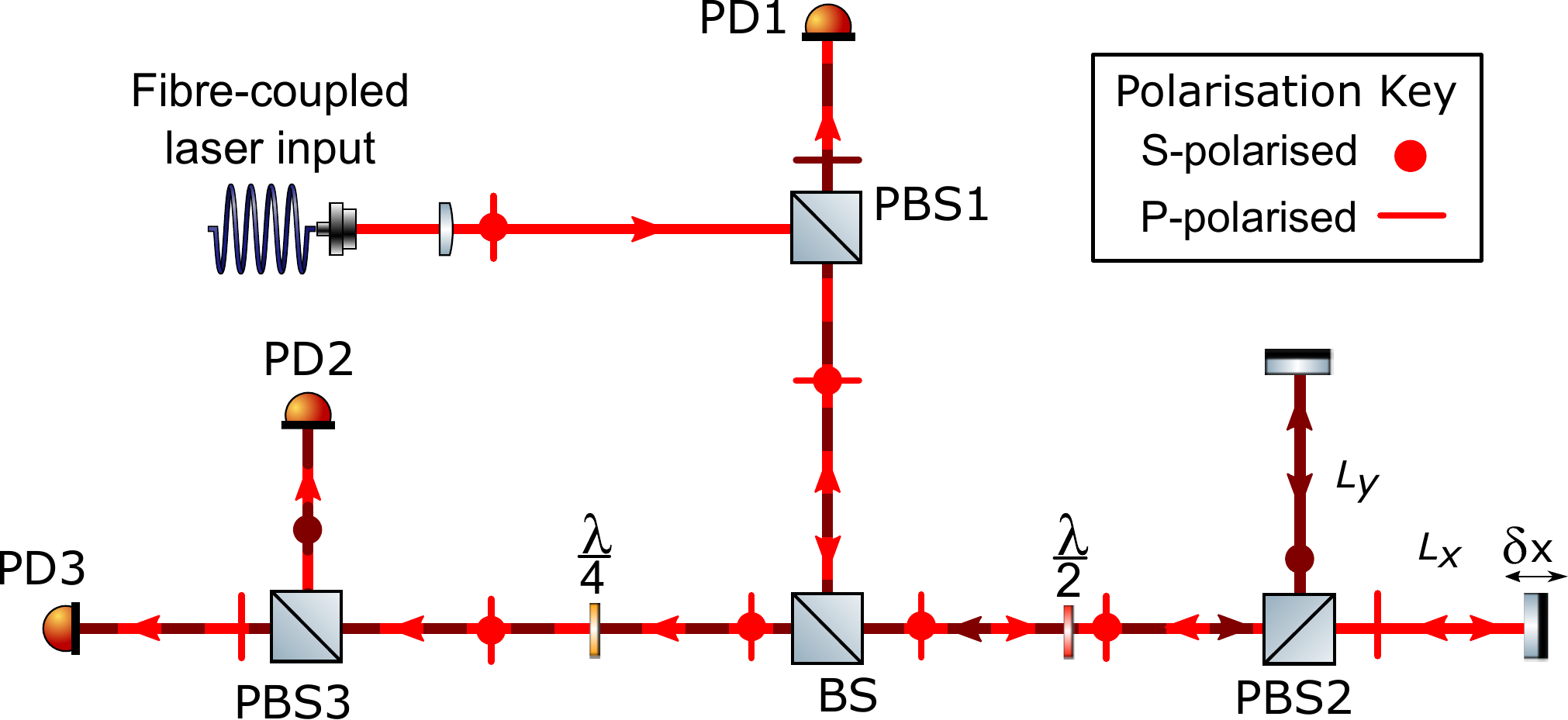}
	\caption{Diagram of the Homodyne Michelson Interferometer with $\lambda/4$ waveplate}
    \label{fig:HoQI_schematic}
\end{figure}

Similarly, a $\lambda/4$ wave plate can also introduce the required phase shift in the system. The phase shift is generated either before entering the two arms~\cite{ponceau08,Aston2011,Bradshaw15,Watchi16} or just before the signals are measured~\cite{Greco95,cooper2017compact}. In the first case, the beam polarization state is rotated before and after entering the interferometer so that both polarizations enter the two arms: one polarization will carry the phase shift $\pi/2$ through the whole optical path. In the second case, after splitting the beam in two thanks to a beam splitter, the phase of one part is delayed by $\pi/2$, see Fig.~\ref{fig:HoQI_schematic}. Here, the first PBS ensures the beam to have a clean polarization state, and the $\lambda/2$ wave plate adjusts this state to ensure that PBS2 splits the beam into two orthogonal polarization states. Note that the configuration in Fig.~\ref{fig:HoQI_schematic} shows more than two photodiodes. The additional photodiode is used to delete the DC component as already explained in Section~\ref{sec:HomodyneExtraphotodiodes}.

Some examples of resolution obtained with the homodyne quadrature interferometers mentioned above are shown in Table~\ref{tab:resolutionEvolutionHomodne}. Even though the use of $\lambda/8$ waveplate eases the optical path, this product is difficult to obtain as this product is not generic. However, the improvements developed for the $\lambda/8$ configuration can be easily implemented on the $\lambda/4$ one. For example, the use of additional photodiodes to delete a DC component presented in Section~\ref{sec:HomodyneExtraphotodiodes} has been used in Ref.~\cite{cooper2017compact} where the $\pi/2$ phase shift is induced by the mean of a $\lambda/4$ waveplate.

\begin{table}[ht]
\centering
\caption{\label{tab:resolutionEvolutionHomodne} Chronological evolution of homodyne quadrature interferometers resolution and other properties. All devices cited uses a waveplate to generate a phase shift of $\pi/2$ between the two polarization states. The resolution is given in Amplitude Spectral Density (ASD) for the interferometers found in the literature and in Root Mean Square (RMS) for the commercial products. The area corresponds to the surface occupied by the interferometer, without the laser source and the data acquisition system.}
\begin{ruledtabular}
\begin{tabular}{cccccc}
\multirow{2}{*}{Year} & \multirow{2}{*}{Device} & Resolution (ASD)& Wavelength & Area \\ 
&  & pm/$\sqrt{\mbox{Hz}}$ @1Hz& nm & cm$^2$ \\
\hline 
2008&Ponceau \cite{ponceau08} & 1& 632.8 & 27x27 \\
2009&Pisani \cite{Pisani09} & 5& 632.8 & 20x20\\ 
2010&Zumberge \cite{Zumberge10}& 0.3 &632.8 & 12x17 \\
2011&Aston \cite{Aston2011}  & 5& 850 & 8.7x4 \\
2012&Acernese \cite{acernese12}& 1 & 632.8 & 13.4x13.4 \\
2015&Bradshaw \cite{Bradshaw15} & 420 & 1550&28x16\\
2016&Watchi \cite{Watchi16}  & 1 & 1550 & 14x11 \\
2017&Cooper \cite{cooper2017compact} & 0.1 & 1064 & 17x10 \\
\hline
\multirow{2}{*}{Year} & Commercial & Resolution & Wavelength & Area \\
&\textcolor{red}{Product} & RMS (pm) & nm & cm$^2$\\
\hline
\multirow{2}{*}{2017} & Renishaw~\cite{Renishaw} & \multirow{2}{*}{38.6} & \multirow{2}{*}{632.8} & \multirow{2}{*}{9.8x5}\\
& RLD10 &&&\\
\multirow{2}{*}{2018} & Zygo~\cite{Zygo} & \multirow{2}{*}{60} & \multirow{2}{*}{633} & \multirow{2}{*}{60x34}\\
& DynaFiz &&&\\
\multirow{2}{*}{2018} &  Dayoptronics~\cite{Dayoptronics} & \multirow{2}{*}{80} & \multirow{2}{*}{632.8} & \multirow{2}{*}{25x12.7}\\
& AK-40 &&&\\
\end{tabular} 
\end{ruledtabular}
\end{table}

\subsubsection{Using a special beam splitter coating}
\label{HomodyneSectionB3}
\begin{figure}[ht!]
	\centering
	\includegraphics[width=1\linewidth]{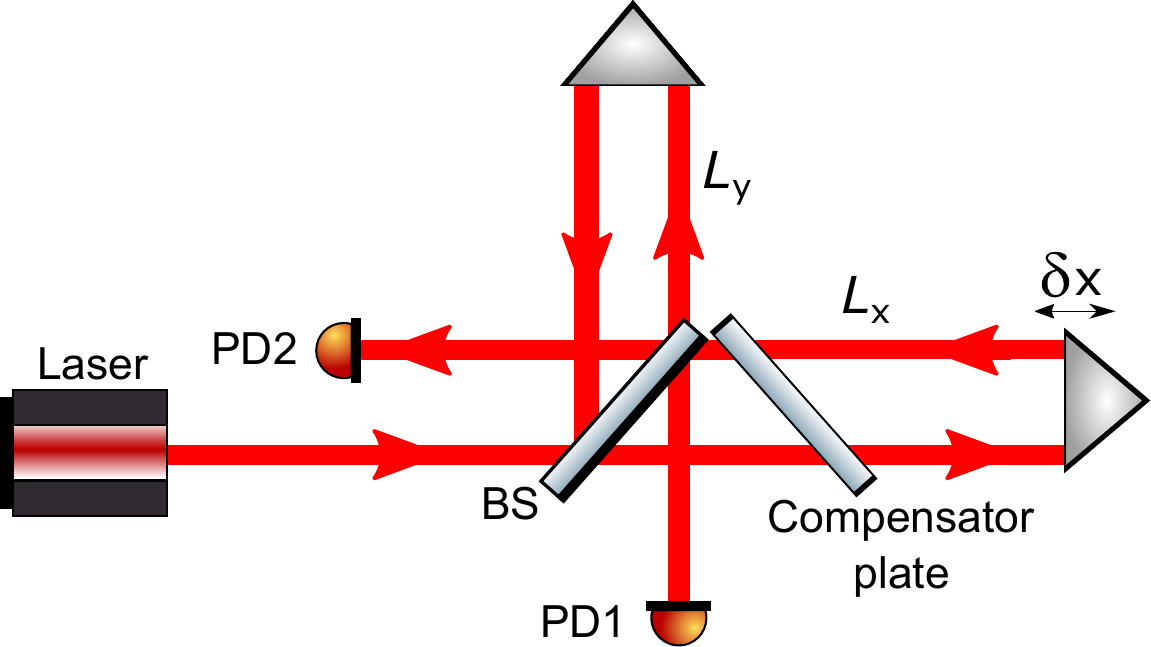}
	\caption{Diagram of the Homodyne Michelson Interferometer with a Special BS Coating}
    \label{fig:coating}
\end{figure}

In order to avoid the unwanted extra reflections that appear when adding wave plates, an interferometer that uses beam splitter plates and corner cubes has been developed~\cite{Downs93}, see  Fig.~\ref{fig:coating}.

In this setup, the BS is replaced by two slightly wedged plates coated with a three-layer metal film~\cite{Downs93}. The beam phase is delayed differently when it is reflected or when it is transmitted through the plates~\cite{Raine78}. With a careful choice of the plate coating, the phase shift between the two path is $\pi/2$ and the two signals are in quadrature. In Ref.~\cite{Raine78}, a method to produce the coating is explained. However, the authors can only guarantee that the phase difference between the two signals is included in the range 90$^{\circ}$ $\pm$ 10$^{\circ}$ which corresponds to a relative uncertainty of more than 10~\%. Consequently, such a beam splitter plate can not provide the phase shift with sufficient precision to ensure that this option can replace the use of wave plates.

\subsection{Quadrature signals carried by transverse electromagnetic modes: tilted mirror}
\label{HomodyneSectionB4}
\begin{figure}[ht!]
	\centering
	\includegraphics[scale=0.8]{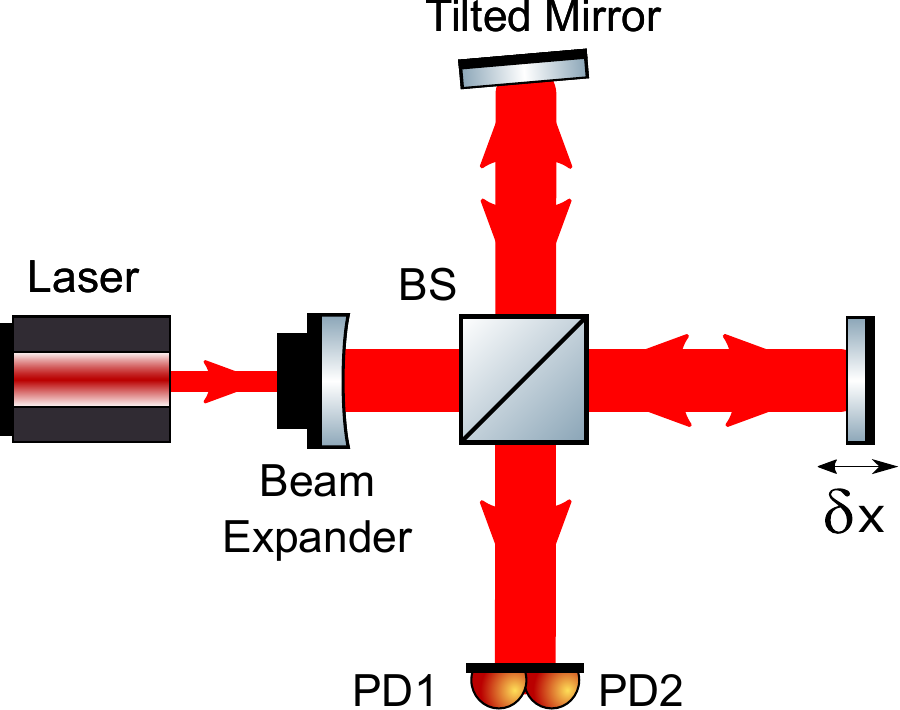}
	\caption{Diagram of the Homodyne Michelson Interferometer with a Tilted Mirror}
    \label{fig:tilted}
\end{figure}

In order to have quadrature signals, the previous method aims to induce a phase shift of $\pi/2$ between the two polarization states of the beam. A phase shift of $\pi/2$ can also be generated between two modes of the intensity beam profile~\cite{slagmolen2002frequency}. In fact, the intensity profile can be seen as a superposition of Transverse Electromagnetic Modes (TEM)~\cite{shaddock2000advanced}. When all optics are well aligned with the cavity of the laser, the intensity distribution of the beam has a Gaussian profile, defined as the $TEM_{00}$ mode~\cite{kogelnik1966laser}. By slightly tilting the mirror of the interferometer, the intensity distribution becomes the sum of a $TEM_{00}$ mode and a $TEM_{01}$ mode. When propagating, these modes accumulate different phase, called a Gouy phase~\cite{shaddock2000advanced,bond2016interferometer}. After travelling, the two modes Gouy phases have aquire a phase shift of $\pi/2$. Consequently, two quadrature signals are measured by placing one photodiode at the maximum intensity of each mode. A diagram of such a device is shown in Fig. \ref{fig:tilted}. The beam expander plays two roles. First, it allows to be in the condition where the phase shift between the two modes is $\pi/2$~\cite{shaddock2000advanced}. Second it eases the positioning of the two photodiodes.

No resolution using this method could be found in the literature. Consequently, its performance will not be discussed.

\section{Heterodyne phasemeters}
\label{sec:heterodyne}

This section will focus on heterodyne phasemeters, this class of interferometers use two frequencies in order to make displacement measurements. The second frequency is generated either by an offset phase-locked laser \cite{Sternkopf12,Hsu10}, through the use of acoustic-optic-modulators (AOM) \cite{Martinussen07,Heinzel03} or using polarization to separate out the beams with different frequencies \cite{Wu2002,Yokoyama99}. There are too many different kinds of heterodyne interferometers to explain them in detail, as such this section will focus on the basic principles behind heterodyne interferometry and go into detail on some specific types of devices.

\textsc{\begin{figure}[ht!]
	\centering
	\includegraphics[width=0.5\textwidth]{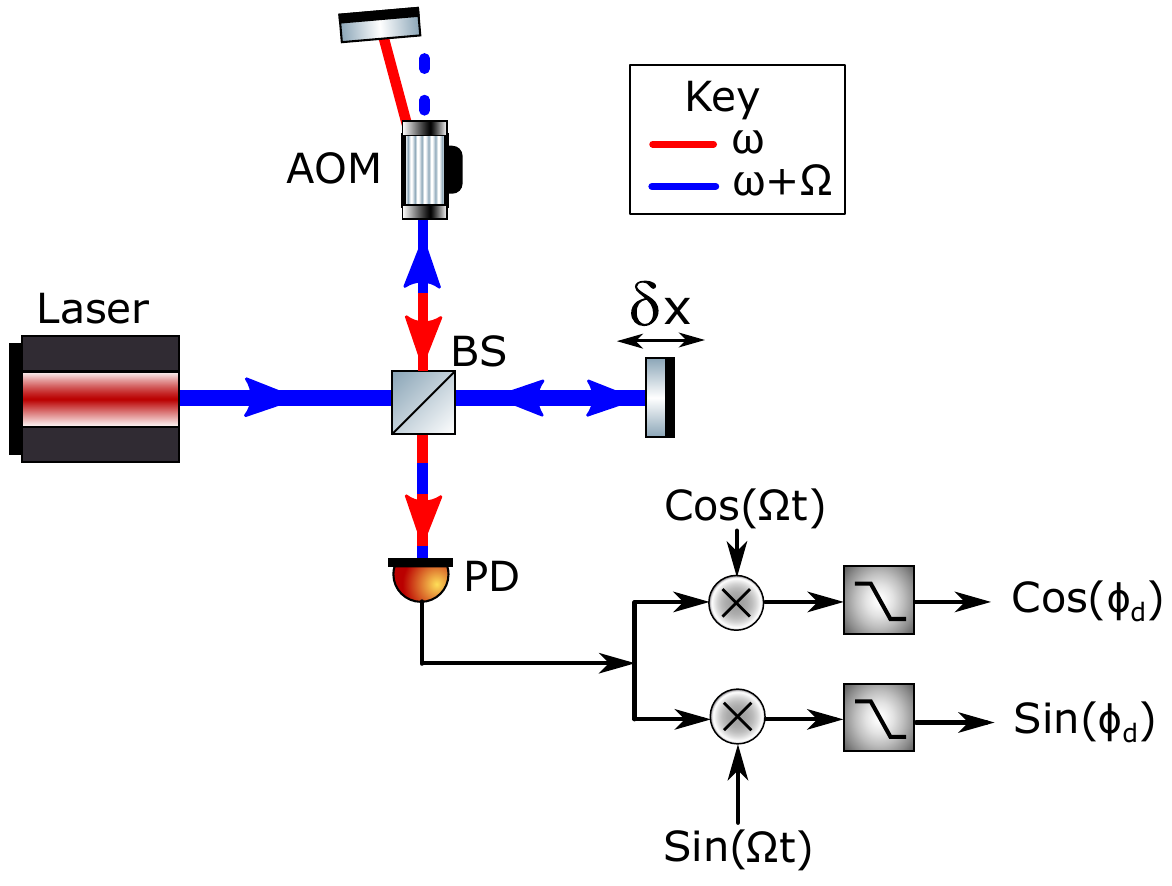}
	\caption{A simple diagram of a heterodyne interferometer.}
	\label{fig:hetIfo}
\end{figure}}

\subsection{Principle of Operation}
\label{sec:heterodyne_principle}
An example of a simple heterodyne interferometer is shown in Fig. \ref{fig:hetIfo}, where a single laser is split into two paths by a non-polarising beam splitter. One of the arms includes an AOM that shifts the laser frequency by \(\Omega\). When the beams interfere, the signal is measured by a photodiode. The two beams acquire phase shifts corresponding to the path lengths of the two arms. However, unlike homodyne interferometers, the beams also pick up an additional phase corresponding to the difference in laser frequency. Assuming an input electric field of \(E_0\), and using the nomenclature found in Section \ref{sec:twobeam}, the output electric field is
\begin{eqnarray}
E_{out} & = & i r t E_0 e^{i \phi_{\rm s}} 
\left(e^{i( \omega t + \Omega t + \phi_{\rm d}/2)} 
+ e^{i (\omega t - \phi_{\rm d}/2)}\right).
\label{eq:hetFieldSimple}
\end{eqnarray}
If we multiply the output electric field by its complex conjugate, the output power is
\begin{eqnarray}
P_{\rm out} & = & \frac{P_{\rm in}}{2} \left(1 +\cos(\Omega t + \phi_{\rm d}) \right).
\label{eq:hetBeatnote}
\end{eqnarray}

It can now be seen that the output contains a beat-note at the difference frequency \(\Omega\), and that the differential phase of the arms, \(\phi_{\rm d}\), is encoded in the phase of this beat-note. 

Unlike homodyne interferometers, heterodyne devices only require a single photodiode in order to measure the differential distance between the two arms across multiple optical fringes, as the beat-note can be demodulated with both sine and cosine local oscillators, this is known as $I/Q$ demodulation. 

For the example in Fig. \ref{fig:hetIfo}, the signal used by the AOM can be used to demodulate the photodiode output signal as shown in Ref.~\cite{rice44}. For the `in-phase', $I$, term we demodulate with a cosine
\begin{eqnarray}
I & = & \frac{P_{\rm in}}{2}\left( 1 + \cos(\Omega t + \phi_{\rm d}) \right)\cos(\Omega t),
\label{eqn:hetACPM} \\ 
& = & \frac{P_{\rm in}}{4}\left(2 \cos(\Omega t) + \cos(2 \Omega t + \phi_{\rm d})  + \cos(\phi_{\rm d})\right),
\label{eqn:hetcos}
\end{eqnarray}
For the `quadrature', $Q$, term, the local oscillator is phase-shifted by 90 degrees, and we demodulate with a sine
\begin{eqnarray}
Q & = & \frac{P_{\rm in}}{2}\left( 1+\cos(\Omega t + \phi_{\rm d}) \right)\sin(\Omega t), \\
& = & \frac{P_{\rm in}}{4}\left(2 \sin(\Omega t) + \sin(2 \Omega t + \phi_{\rm d}) - \sin(\phi_{\rm d}) \right).
\label{eqn:hetsine}
\end{eqnarray}
Terms with multiples of \(\Omega t\) are removed with an appropriate low-pass filter, leaving only the final term containing the optical phase. The frequency of the low pass filter and the heterodyne frequency are interlinked, the value of the former effectively sets the latter's frequency. The optical phase can then be unambiguously extracted over many wavelengths by (unwrapping) the output of a 4-quadrature arctangent, as in the homodyne phase meter case as shown in section \ref{sec:homodyne}. 

Heterodyne interferometers are sometimes classed as AC phasemeters, as the differential optical phase is encoded in the phase of the beat frequency between the two lasers, as shown in Eq. \ref{eq:hetBeatnote}. This allows the interference to be measured at the modulation frequency, typically in the KHz to MHz region, away from low-frequency noise sources that may couple into the measurement \cite{Hechenblaikner13}, such as laser intensity noise and electronic `$1/f$' noise, improving low-frequency performance. 

The drawback is that they use additional optical and opto-electronic components to generate the second frequency, typically resulting in increased complexity, expense and size compared with homodyne phasemeters. Moreover a suitable lowpass filter needs to be chosen in accordance with the demodulation frequency. These must be chosen appropriately to remove the high frequency beat terms while still allowing the optical phase to be read out unattenuated. 

\subsection{Comparison of Devices}
\subsubsection{Single Photodiode Devices}
Early heterodyne interferometers, as the one shown in Fig.~\ref{fig:hetIfo}, operated in the MHz-GHz frequency band, but over time similar levels of resolution and reductions in non-linearity have been achieved using lower modulation frequencies. In 1970, HP released a commercial heterodyne interferometer boasting an accuracy of $10^{-8}$~m, running at a modulation frequency of a 2\,MHz \cite{hpinterferometer70}. In De la Rue \textit{et al.} \cite{delaRue72} they employ a Bragg cell heterodyne interferometer to measure acoustic waves, and they achieve a resolution of 0.2\,pm / \(\sqrt{\rm{Hz}}\) at 2\,Mhz. Monachalin \cite{Monchalin84} uses a similar optical layout but with a commercial available lock-in amplifier and achieves a detection limit of 60\,fm / \(\sqrt{\rm{Hz}}\) at 1\,Hz.  Royer and Dieulesaint \cite{royer86} improve on this resolution and present a compact (8x5x3\,cm) heterodyne interferometer, with a peak resolution of 30\,fm / \(\sqrt{\rm{Hz}}\) at 1\,Hz. This device offers improvements in terms of ease of alignment and improved stability due to compactness of the optics. 

Martinussen \textit{et al.} \cite{Martinussen07} present a heterodyne interferometer with pico-meter resolution operating in the 0-1.2\,GHz regime to measure properties of capacitive micro-machined ultrasonic transducers and has peak resolution of 4\,pm in this range. Here the second laser frequency is generated in the reference arm of the interferometer. Leirset \textit{et al.} \cite{Leirset13} improve upon this design by focusing the beam on the input of the AOM and report a significant resolution improvement of 7.1\,fm / $\sqrt{\rm{Hz}}$ at 21\,Mhz.
In Willemin \textit{et al.} \cite{Willemin87} a heterodyne interferometer is proposed to measure vibrations in the inner ear, the device used in these experiments has a resolution 30\,pm  / \(\sqrt{\rm{Hz}}\) at 1\,Hz. 

\subsubsection{Reference Photodiode Devices}
\label{sec:ReferencePhotodiodeDevices}
\begin{figure}[ht!]
	\centering
	\includegraphics[width=1\linewidth]{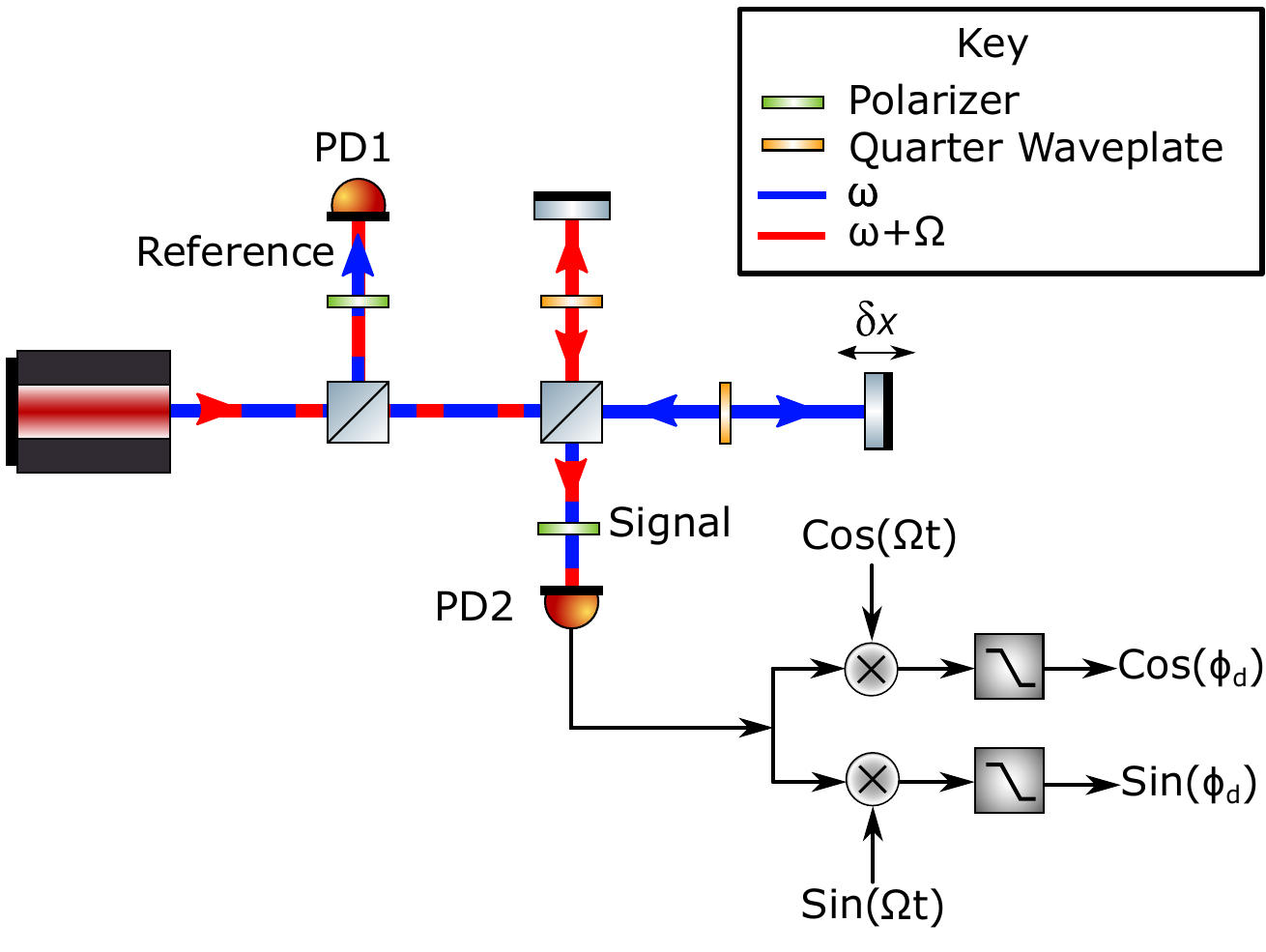}
	\caption{The readout scheme of a heterodyne phasemeter with two different laser frequencies which are spatially separated, adapted from Ref.~\cite{Bradshaw15}}
	\label{fig:heterodynephasemeter}
\end{figure}

Heterodyne phasemeters can achieve exceptional resolution at lower frequencies than those presented above by using a reference photodiode, in addition to the signal photodiodes, to increase common-mode rejection. Polarization or frequency shifts that occur in the interferometer, but outside the measurement arms, can be measured and cancelled. This can be achieved by de-modulating the signal photodiode (PD2 in Fig. \ref{fig:heterodynephasemeter}) with the output of a reference photodiode (PD1 in Fig. \ref{fig:heterodynephasemeter}), suppressing common fluctuations in the base-band output \cite{Gasvik03}. An alternative and equivalent method is to independently extract the phase of the light on the two photodiodes and subtract them. These two approaches provide the same result, although the second is conceptually simpler. 

\begin{figure}[ht!]
	\centering
	\includegraphics[width=1\linewidth]{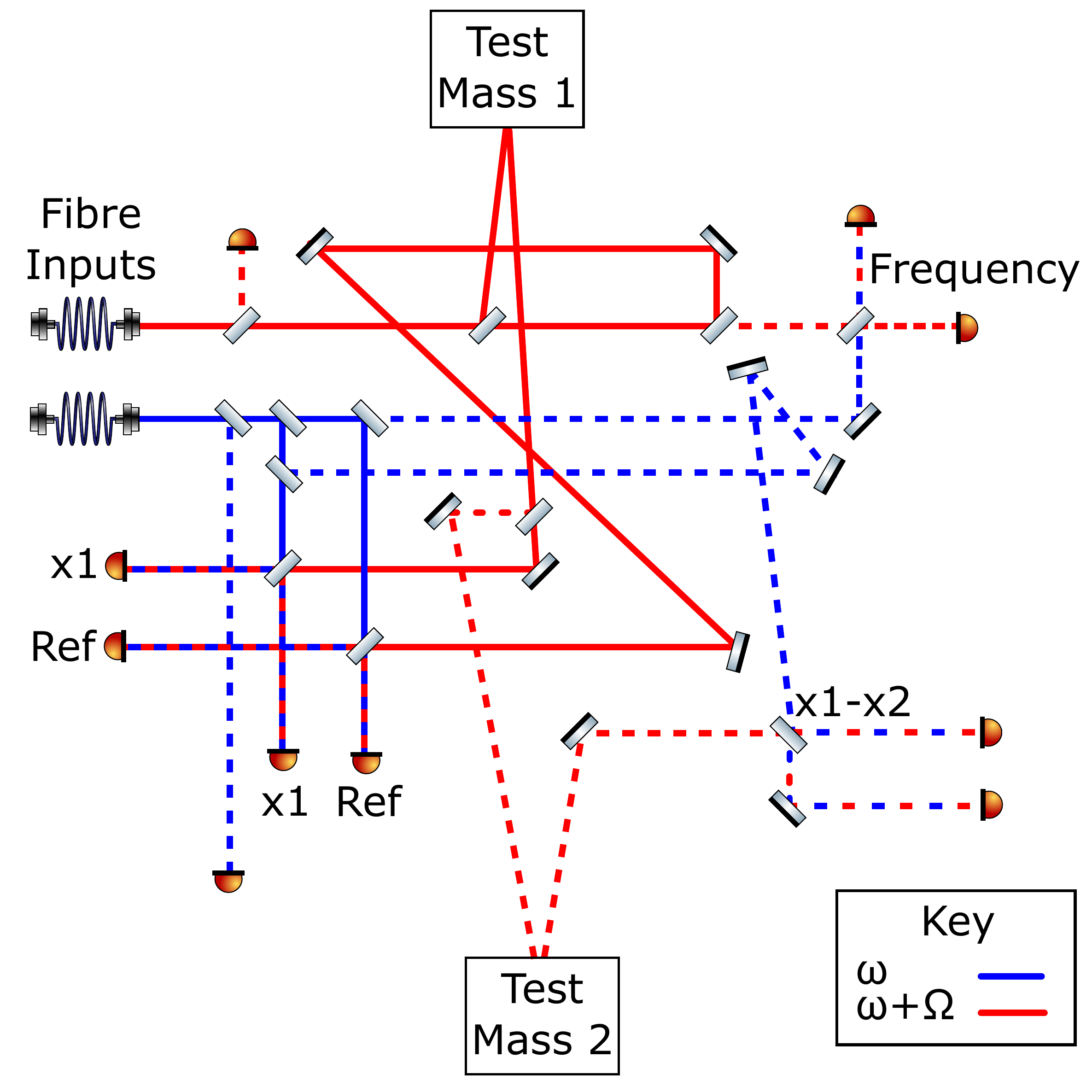}
	\caption{LISA Pathfinder employs four independent heterodyne phasemeters. The two frequencies used to generate the beat-note are represented in red and blue. The `reference' and `x1' paths are highlighted by solid lines, other phasemeters are shown with dashed lines, adapted from Ref.~\cite{Heinzel03}}
	\label{fig:LPFHET}
\end{figure}

The LISA Technology Pathfinder's (LPF) optical readout, seen in Fig. \ref{fig:LPFHET}, employs the second technique. The power on the reference photodiode has a beat-note at the difference frequency and a time-fluctuating phase that is common to all interferometers, $\phi_{\rm c}$, due to fluctuations on the input beams
\begin{eqnarray}
P_{\rm {ref}} & \propto & P_{\rm{in}} (1 + \cos(\Omega t + \phi_{\rm{c}}))
\end{eqnarray}

If we then follow the path corresponding to the interferometer that measures position of the first test mass, the X1 photodiode, the signal measured is simply:

\begin{eqnarray}
P & = & \frac{P_{\rm{in}}}{8}(1+\cos(\Omega t + \phi_{\rm c} + kL_1))
\end{eqnarray}

Where \(P_{in}\) and \(\omega\) are as before, $k$ is the wave number and \(L_1\) is the path length between the optical board and the test mass. Once the two phases have been extracted using the technique described previously the common phase between the two paths can be subtracted, leaving the optical phase caused by the motion of the test mass. We find the two optical phases and the resultant phase as follows:

\begin{eqnarray}
\phi_{\rm{ref}} & = & \Omega t + \phi_{\rm c} \\
\phi_{\rm{x1}} & = & \Omega t + \phi_{\rm c} + kL_1 \\
\phi_{\rm{x1}} - \phi_{\rm{ref}} & = & kL_1
\end{eqnarray}


Schuldt \textit{et al.} \cite{Schuldt09} report on a heterodyne interferometer designed as a demonstrator for proof mass translation onboard the LISA satellites inside a vacuum chamber. With intensity stabilisation, the  interferometer achieves a resolution of 10\,pm / \(\sqrt{\rm{Hz}}\) and 2\,pm / \(\sqrt{\rm{Hz}}\) at 10\,mHz and 1\,Hz respectively.  

LISA pathfinder is a space based mission to test technology prior to the launch of the LISA gravitational wave detector, currently stated for launch in the 2030's. This gravitational wave detector aims to detect astrophysical events at very low frequencies from 1\,Hz down to 0.1\,mHz. The optical bench interferometer is comprised of four heterodyne interferometers \cite{Robertson13}, these operate over a large dynamic range by using reference interferometer to provide a main phase reference. This common phase reference suppress low frequency noise sources such as thermal expansion of the optical fibres and noise due to the AOM driver noise. Unlike the optical configurations presented previously, the interferometers in LISA pathfinder do not use polarization optics to separate out the reference and signal beams as these may have induced too much low frequency noise into the interferometer \cite{Heinzel03}.  Complete details of the optical setup are described in Heinzel \textit{et al.} \cite{Heinzel04,Heinzel05,steier09}. The phasemeter on-board LISA pathfinder has a resolution of 1\,pm / \(\sqrt{\rm{Hz}}\) at 10mHz \cite{Armano16}.
\subsubsection{Polarization Based Devices}
Devices such as those described in Refs.~\cite{Wu2002,Zhao16} employ polarization optics, frequency and spatially separated beams, see Fig.~\ref{fig:heterodynephasemeter}. In this configuration the second laser frequency is spatially separated from the main laser frequency and thus doesn't interact with the reference or target mirrors. This spatial seperation is said to avoid non-linearities caused by polarization and frequency mixing.  A reference beam is used to track the heterodyne beat-note. The signal at the measurement photodiode, PD2 in Fig. \ref{fig:heterodynephasemeter} can then be demodulated with the signal at PD1, effectively subtracting the common phase. The reported resolution is 2 and 5\,pm  / \(\sqrt{\rm{Hz}}\) at 1.7 and 5\,KHz respectively in Ref.~\cite{Wu2002}  and 3\,pm  / \(\sqrt{\rm{Hz}}\) at 45\,Hz in Ref.~\cite{Zhao16}.

Hsu \textit{et al.} \cite{Hsu10} present a Sagnac interferometer, employing both polarization and a reference photodiode to reach a peak resolution of 0.5\,pm/ \(\sqrt{\rm{Hz}}\) above 10\,Hz. The input beam is split in two, counter propagate and pick up phase shifts due to the 2 AOM's in the arms of the interferometer. As well as its impressive resolution, the interferometer also achieves more than 70\,dB of common-mode noise suppression.

In the past ten years, a major focus has been in the reduction of non-linearities in the readout as shown by Weichert \textit{et al.} and Pisani \textit{et al.} \cite{weichert2011,pisani12,Pisani09}. These devices are based off a single, frequency stabilised laser that is locked to a hyperfine structure line in Iodine. The device contains two AOM's, producing two separate frequency shifted beams at 78.4375 and 80\,Mhz respectively. These two beams are split once more forming two interferometers, one using the 80\,MHz beam in the reference arm, with the 78.4375\,MHz beam being used in the signal arm. In the second interferometer, the roles of these beams are reversed. This method means that each sub-interferometer can be considered to be independent of the other, allowing drift in the AOM driving frequency, that would otherwise couple into the readout to be eliminated \cite{weichert12}.

Careful attention in this configuration was made to ensure the beat frequency was sufficiently higher than the resonant frequencies of the input fibres and that the laser source was spatially separated from the interferometer to minimise thermal noise coupling. This configuration achieves a resolution of 0.03\,pm / \(\sqrt{\rm{Hz}}\) at 1\,kHz..
\subsubsection{Deep Phase Modulation}
\label{sec:PhaseModulation}
In order to work over many fringes, the beam frequency can also be modulated by a sinusoidal phase. This sinusoidal phase can be applied by two different ways to an interferometer. First, assuming that in one part of the arms, the beam propagates in an optical fiber, a piezo is applying a sinusoidal motion to the optical fibre \cite{Teran15}. Second, an electro-optical amplitude modulator is modifying the laser frequency with a sinusoidal signal \cite{Isleif16,gerberding15}. The first method is called Deep Phase Modulation (DPM) while the second one is called Deep Frequency Modulation (DFM). The phase is extracted similarly as for the other types of heterodyne interferometers, see section~\ref{sec:heterodyne_principle}: the in-phase and quadrature terms are evaluated then low-pass filtered to extract the phase.

For both cases, the methods are combined with a non zero optical path length difference interferometer and an appropriate demodulation algorithm which can be implemented on a FPGA \cite{Teran15}. Miniaturisation of the device is thus possible.

In Ref.~\cite{gerberding15}, it has been proven that DFM suppress fiber length noise. The remaining dominant noise source which can not be suppressed is the laser frequency noise. To reduce this noise source, the laser is injected into a stable, unequal arm length interferometer. In this reference interferometer the laser frequency noise can be measured, and the associated optical phase can be subtracted from the DFM interferometer \cite{gerberding15}.
 
\begin{table*}
	\centering
	\caption{\label{tab:hetSummary} Chronological evolution of the resolution of heterodyne interferometers. The area corresponds to the surface occupied by the interferometer, without the laser source and the data acquisition system.}
	\begin{ruledtabular}
	\begin{tabular}{ccccccc}
		Year & Device & Resolution (ASD) & Resolution Meas-& Heterodyne  & Wavelength & Area\\ 
		& & pm / \(\sqrt{\rm{Hz}}\) & urement Frequency & Frequency & nm & cm$^2$\\
		\hline 
		1970  & HP \cite{hpinterferometer70} & 10,000 & - & 2\,MHz & 632 & 38x28\\ 
		\hline 
		1972  & De la Rue \cite{delaRue72} & 0.2  & 2\,Mhz & 22.5\,MHz & 632 & -\\ 
		\hline 
		1984  & Monachalin \cite{Monchalin84} & 0.06 & 1\,Hz & 40\,MHz & 632 & - \\ 
		\hline 
		1986  & Royer \cite{royer86} & 0.1 & $>$100\,kHz & 70\,MHz & 632 & 8x5\footnotemark[1] \\ 	
		\hline 
		1987  & Willemin \cite{Willemin87} & 10 & 1\,kHz & 1\,MHz & 632 & - \\ 
		\hline
		2002  & Wu \cite{Wu2002} & 2 & 1.7\,kHz & 80\,kHz & 632 & 40x40\footnotemark[1]  \\ 
		\hline 
		2007  & Martinussen \cite{Martinussen07} & 2 & 3.3\,Hz & 31\,MHz & 532 & -\\ 
		\hline 
		2009  & Schuldt \cite{Schuldt09} & 10 & 0.01\,Hz & 10\,kHz & 1064 & 30x40\\ 
		\hline  
		2010  & Hsu \cite{Hsu10} & 0.5 & 10\,Hz & 1.65\,MHz & 632 & - \\ 
		\hline
		2012  & Weichert, Pisani \cite{weichert12,Pisani15}  & 0.03 & 1\,kHz & 1.5625\,MHz& 532 & -  \\ 
		\hline	
		2013 & Leirset \cite{Leirset13}  & 0.071 & 21\,MHz & 0-1.3\,GHz & 532 & - \\ 
		\hline 
		2016  & LPF \cite{Armano16} & 1  & 10\,mHz & - & 1064 & 20x20\footnotemark[1]  \\  
	\end{tabular} 
	\end{ruledtabular}
	\footnotetext[1]{Interferometer plate only}
\end{table*}

\subsection{Conclusion}
While operating at the beat-note is an advantage in terms of simplicity of the readout, interferometers that use this technique are still subject to the same fundamental noise sources. These include but aren't limited to: shot noise and length noise coupling into the readout. The effective contribution of laser frequency noise can be reduced as shown in Ref.~\cite{Heinzel03} and Ref.~\cite{Wu2002}.
In terms of low frequency resolution, the phasemeter on board of the LISA pathfinder spacecraft represents the best heterodyne phasemeter in terms of linearity and sensitivities below 1\,Hz, however the interferometer is expensive when compared to other devices. The most compact interferometer reviewed here, with a specified size is developed by Royer \textit{et al.}\cite{royer86}, this device has excellent resolution of 30\,fm / \(\sqrt{\rm{Hz}}\) at 70\,MHz, however the device does not specify its linearity. The most linear interferometer is that presented by Weichert \textit{et al.} \cite{weichert12}  with non-linearities less than 5\,pm and a noise floor of 30\,fm / \(\sqrt{\rm{Hz}}\) above 150\,Hz, though it lacks the simplicity of devices such as the one presented in Leirset \textit{et al.} \cite{Leirset13}. A summary of the interferometers reviewed and their subsequent sensitivities are shown in chronological order in Table \ref{tab:hetSummary}.

\section{Linearity of Phasemeters}
\label{sec:nonlinearities}

Phasemeters recover the optical phase by evaluating the four-quadrant arctangent of the ratio between two quadrature signals. The relation between the real phase and the phase measured should be linear but there are often distortions due to spurious effects in the optics or signal-processing of the phase.  These distortions correspond to non-linearities and cause periodic errors of the relation between the real phase and the measured phase. Techniques to reduce and quantify non-linearities, are the scope of this section.

\begin{figure}[ht!]
	\centering
	\includegraphics[trim = {0cm 0cm 0cm 0cm},width=1\linewidth]{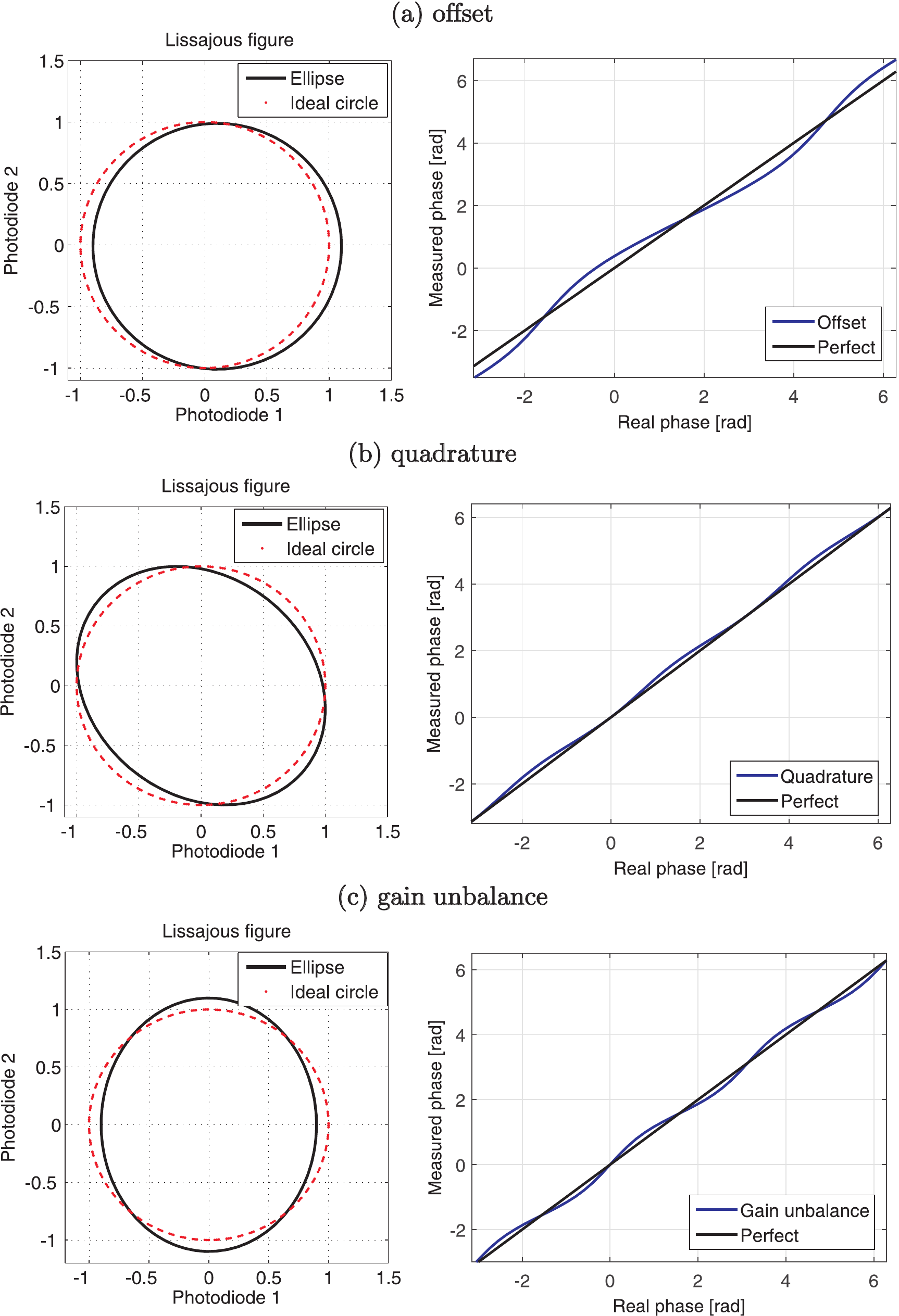}
	\caption{Plots of $P_1$ against $P_2$ (left) and the effect of the non-linearity on the relation between the real and the measured phases (right). The effect of offset (a), quadrature error (b), and gain imbalance (c) can be seen in the Lissajous figures when compared with an ideal circle. For simplification, the circles and ellipses are centred at the origin. The right figures allow to identify the order of the non-linearity in comparison to the period of the sinusoidal signals: the offset has an order~1 and the quadrature and the gain imbalance an order~2.}
    \label{fig:NLcombined}
\end{figure}

The ideal signal of a homodyne interferometer is a sinusoidal shape Eq.~\ref{eq:mich}. For a quadrature homodyne interferometer, it is a circular Lissajous figure Eq.~\ref{eq:iqSimple}. These perfect patterns are distorted by offset Fig. \ref{fig:NLcombined}(a), quadrature imperfections Fig. \ref{fig:NLcombined}(b), and gain imbalance of the signal due to an intensity difference between the two arms of the interferometer Fig. \ref{fig:NLcombined}(c). The resulting Lissajous figure is a rotated ellipse. The phase recovered from this figure is different from the real phase~\cite{Wu2002} 
and the signals measured for a homodyne interferometer have the following form~\cite{Otero09}:
\begin{eqnarray}
P_1 &=& P_0(1 + a \cos(\phi_{\rm d})) \label{eq:ellipseX}\\
P_2 &=& b \, P_0 (1 + a\sin(\phi_{\rm d} + c)) + d  \label{eq:ellipseY}
\end{eqnarray}
where $P_1$ and $P_2$ are the measured signals as in Eq. \ref{eq:iqSimple}, $P_0$ is proportional to the laser power, $a$ is the fringe visibility, $b$ is the gain mismatch between sensors, $c$ is the quadrature imperfection, and $d$ is the differential offset.

Some heterodyne interferometers have the same non-linear behaviour such as the one described in Fig.~\ref{fig:hetIfo}. However, the optical configurations like the one in Fig.~\ref{fig:heterodynephasemeter} encounter other sources of non-linearities, mainly due to phase mixing in the two arms of the interferometer. A complete description of this last type of distortion can be found in Ref.~\cite{Xu13}. This section will more focus on the non-linearities engendered in homodyne-like interferometers. %

As seen in Fig.~\ref{fig:NLcombined}, distortions due to translation and dilatation of the Lissajous figure induce a periodic variation of overestimation and underestimation of the phase. In fact, over one period, the sine and/or cosine are alternatively smaller and bigger than the ideal case. On the contrary, the rotation of the figure corresponds to an additional constant phase applied to one of the two signals. Depending on the phase sign, this extra phase is responsible of either an overestimation or an underestimation of the relation between the real and measured phases.

Causes of non-linearities, include:
\begin{itemize}
\item Elliptical polarization of the laser beam \cite{Xu13,Bobroff93}
\item Misalignment between the laser beam and the beam splitter polarization axis  \cite{Hou92,Rosembluth90}
\item Imperfections in alignment or quality of optical components \cite{Hou92,Rosembluth90}
\item Non-orthogonality of the laser polarizations \cite{Hou92,Ju15,Petru99}
\item Imperfect photodiode (responsivity, gain, etc.)~\cite{Ellis14}
\end{itemize}

This non-exhaustive list shows the complexity of the non-linear origins \cite{Rosembluth90}. Moreover, one cause of non-linearity engenders combinations of offset, quadrature and gain imbalance distorsions. For example, if the two polarization states are not perfectly orthogonal, the two polarizations measured will not have the same intensity and they will not be in quadrature. 

In order to reduce the sources of non-linearities, several solutions have been implemented: ellipse fitting algorithms, phase-lock systems, temperature isolation, etc. The different techniques and the improvements brought are listed below. The corresponding residual non-linearities are gathered in Table~\ref{tab:nonLinearities}.

\subsection{Ellipse fitting algorithms}
\label{sec:EllipseFittingAlgorithm}
In order to convert the ellipse into a unitary circle, the ellipse parameters in Eq.~\eqref{eq:ellipseX} and \eqref{eq:ellipseY} need to be determined. This can be done by using ellipse fitting algorithms either in post-processing or real-time. 
Algorithms that employ the method of least squares have been used to reconstruct the ellipse parameters~\cite{Rosin93,watkins14} and then recover the parameters from Eqs.~\ref{eq:ellipseX} and~\ref{eq:ellipseY}~\cite{Heydemann81, wu96, Zumberge04, Otero09, Gregorcic09, Hou92, Pozar11, Petru99, Koning14}. In Ref.~\cite{Goldberg01}, the phase error is compensated in the Fourier domain by a least squares approximation of the first order errors. A clear explanation of this ellipse fitting technique is contained in Ref.~\cite{Rosin93}.

In order to identify the ellipse parameters, a cost term, $S$, is minimised. Using the algebraic distance between data and fit points $Q(x,y)$ 
\begin{eqnarray}
S &=& \sum^n_{i=1} Q(x_i,y_i)^2,
\end{eqnarray}

\begin{figure}[ht!]
    \centering
	\includegraphics[trim ={0.4cm 0.7cm 0.8cm 0.5cm},width=1\linewidth]{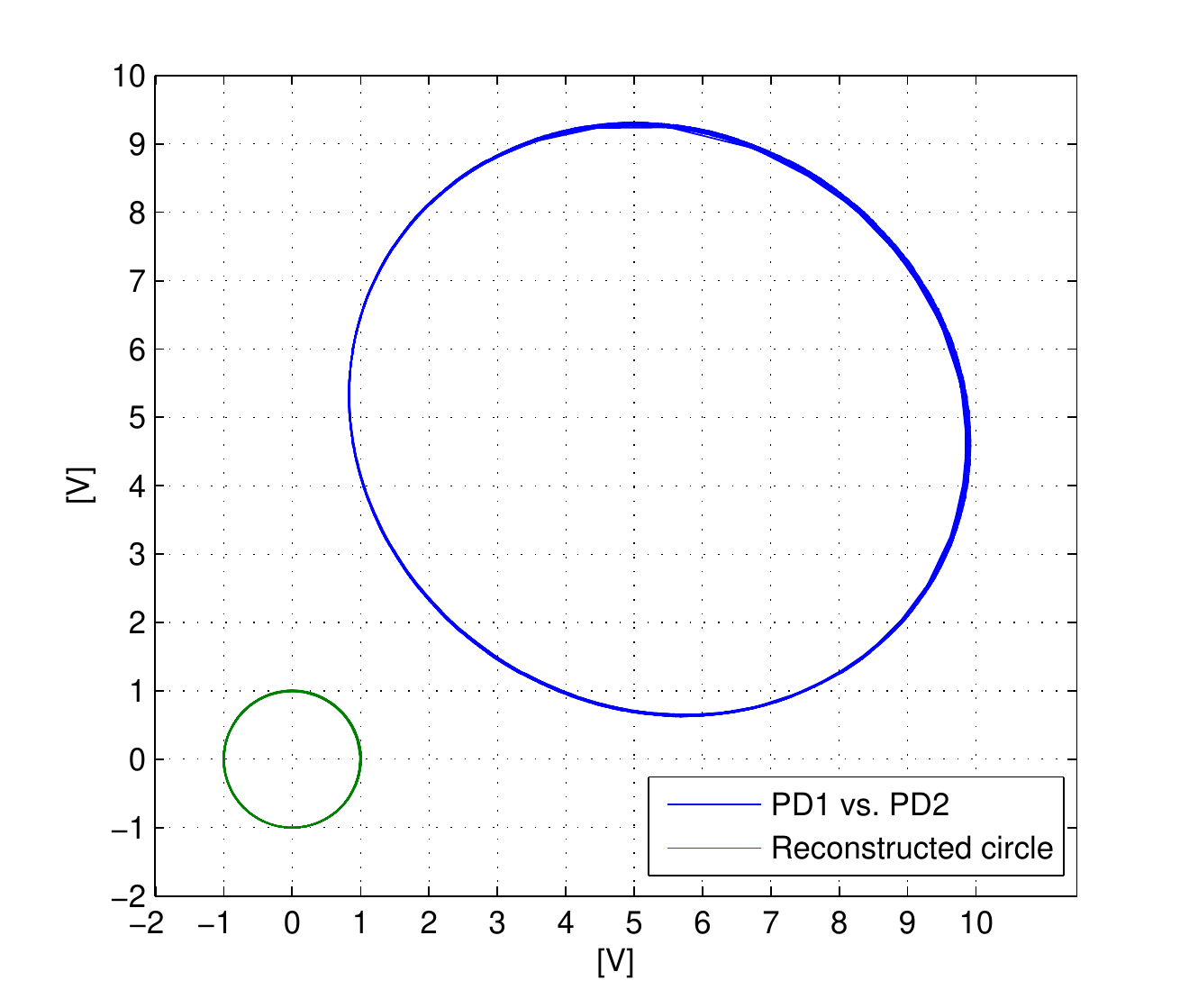}
\caption{Transformation of the ellipse, the signal directly measured by the two photodiodes PD1 and PD2 (blue curve), into a unitary circle (green curve) using ellipse fitting algorithm~\cite{Watchi16}. \label{f:unitaryCircle}}
\end{figure}

In Fig.~\ref{f:unitaryCircle}, the reconstrution of circle thanks to ellipse fitting algorithm is illustrated on  experimental data~\cite{Watchi16}.

In these algorithms, some parameters need to be correctly chosen in order to reduce the non-linearities. First, the fit point on the ellipse closest to the data point has to be properly chosen~\cite{Collett14}. Second, least squares method is very often used and the residual non-linearities with this fitting method are on average between 0.1 and 1~nm, see Table \ref{tab:nonLinearities}. However, other fitting methods exist which reduce the non-linearities. In Ref.~\cite{Emancipator93}, the phase is fitted by a polynomial function and in Ref.~\cite{Collett15}, the parameters are dynamically re-evaluated by iterative refinement. An iterative evaluation is also presented in Ref.~\cite{wang17} where Kalman filters are used to estimate the ellipse parameters. Moreover, the size and shape of the window sampling function used is a crucial parameter for the algorithm performance. The influence of the window function on the phase error has already been studied theorically and experimentally~\cite{deGroot95,Schmit96}: rectangular windows are more sensitive to high-frequency phase errors than bell-shape windows like Von Hann~\cite{deGroot95} and Hanning~\cite{Schmit96} windows. 

\subsection{Non-linearities reduction methods}
Correcting the signal measured is not the only mean to reduce non-linearities. Modifications of the optical path in the interferometer can also improve the signal. Proposed solutions and their performance are discussed in the next sections.

\subsubsection{Multiple reflection in the measurement arm}
\label{sec:MultipassNL}
In section~\ref{sec:MultipassHomodyne}, it has been shown that the multiple reflection technique improves the resolution of homodyne interferometer of a factor $G$~\cite{Pisani09}. With this configuration, the distortions on the resulting signals are similar to the ones obtained with a simple homodyne interferometer. However, as the signal has travelled a longer distance, it has crossed more fringes. From Fig.~\ref{fig:NLcombined}, we can see that non-linearities are periodic and do not increase depending on the number of fringes crossed. Consequently, the ratio between the non-linearities and the whole signal is reduced of a factor $G$ in a multiple reflection interferometer. However, this assumption has not been verified experimentally. 

\subsubsection{Additional sensors}
\label{sec:AdditionalSensors}
As the laser intensity fluctuates, the use of one~\cite{ponceau08,daniel2005advanced} or two~\cite{Greco95,Eom01} additional signals to normalize the measurements reduces the gain imbalance, seen in Fig. \ref{fig:NLcombined}(c). In term of accuracy, two additional photodiodes is more effective because it does not require additional modelling to reduce all types of non-linearities, as explained in Ref.~\cite{Greco95}.

The four photodiodes design can recover in real time all the ellipse parameters of Eq.~\eqref{eq:ellipseX} and Eq.~\eqref{eq:ellipseY} thanks to an electronic circuit. To obtain four signals, two for each polarization state, two PBS are used. This technique reduces all the major types of non-linearity, and the resulting signal had phase error reduced by a factor 10~\cite{Eom01}.

One additional photodiode can be used in two different ways to cancel or reduce intensity fluctuation and offset. In Ref.~\cite{ponceau08}, the additional signal is used to monitor the input power and normalise the outputs from the signal photodiodes. This makes the two signals independent of intensity fluctuations and in a secondary way this reduces gain imbalance. Moreover, the offset non-linearity is reduced as the signal is divided to make the normalisation. 
\\
In Ref.~\cite{Downs79,cooper2017compact}, two signals measured are out of phase and one signal is in quadrature with the two others as already explained in Section~\ref{sec:HomodyneExtraphotodiodes}. If we don't consider a gain mismatch between sensors, see Eq.~\ref{eq:ellipseX} and Eq.~\ref{eq:ellipseY}, quadrature imperfection and a differential offset (as they are not altered with this method), Eq.~\ref{eq:P1_3PD} and Eq.~\ref{eq:P2_3PD} become
\begin{eqnarray}
P_1 &=& \sqrt{2}aP_0 \sin(\phi_{\rm d}-\frac{\pi}{4})\\
P_2 &=& \sqrt{2}aP_0 \sin(\phi_{\rm d}+\frac{\pi}{4})
\end{eqnarray}
As the phase is obtained from the atan2 of the ratio between these two signals, the results become insensitive to the input power fluctuations, which is the parameter $a$ in these equations.

\subsubsection{Reduction of the phase mixing}
\label{NonlinearitiesSectionC}
\label{sec:PhaseMixing}
In homodyne interferometers, when a fraction of one polarization state propagates in the other interferometer arm, we talk about phase mixing. In fact, both polarizations states will then carry information about the reference and measurement arm as they have propagated in both arms. Heterodyne interferometers are also subjected to phase mixing when one of the two frequencies is transmitted to the other path. 
This phase mixing is responsible of imperfect quadrature and gain imbalance as shown in Ref.~\cite{Xu13}.  
Phase mixing can come from imperfect optical elements~\cite{Ju15} such as PBS or optical fibers.

In order to avoid the injection of one polarization state (or wavelength for the heterodyne interferometer) into the other arm, one solution is to make the signals travel into two spatially separated paths and measure the signals with two independent photodiodes. One example of spatial separation can be found in Ref.~\cite{Pisani15}: the central part of the beam cross section is reflected by the measuring mirror and measured by one photodiode. The outer part of the beam is reflected by the reference mirror and recorded by a second photodiode. Note that with this configuration, some diffraction at the separating optics can cause injection of one phase into the other arm but this effect can be reduced thanks to a careful sizing of the setup~\cite{Pisani15}.
\\
Spatial separation is also implemented in Ref.~\cite{Wu2002,weichert12} where two lasers with different frequency propagate in two different interferometers: the only common element between the two interferometers is the moving mirror but the beams are not reflected at the same position on the mirror. With this configuration, interference occurs at the photodiodes where the two beams recombine. 

\subsubsection{Actuators to decrease the non-linearities}
\paragraph{Frequency correction}
It is well known that laser frequency oscillates around a fixed value. This fluctuation creates some phase shift that can be misinterpreted as being a displacement signal. To avoid these fluctuations, the frequency of some interferometers lasers is locked by the mean of a phase-lock system: a reference signal, measured before the beam enters the interferometer, is used to drive the laser cavity. This method is used for homodyne interferometers~\cite{Teran15}, heterodyne interferometers~\cite{Xu13,wu13,Robertson13,gerberding15} and resonators~\cite{Celik12,Yacoot09,pisani12,Guzman14,Guzman15OMSensor,attocube09}.  
Some papers discuss the implementation of frequency lock techniques~\cite{Guzman14,Guzman15OMSensor,gerberding15,Teran15}.

The disadvantage of actuating the laser frequency is that the non-linearity of the measured signal is transmitted to the actuator which will then have a non-linear behaviour. However, as already mentioned, this paper is focussed on wide range readout and not on wide range closed loop readout. The transmission of the non-linearities to the actuator will thus not be further discussed.

Note that the undesired frequencies can be rejected without any control. In Ref.~\cite{gregorvcivc2009quadrature}, an optical narrow band pass filter is placed before the photodiodes. This filter reduces beam signals which do not have the desired wavelength.

\paragraph{Polarization correction}
Misalignment between the polarization states of the incoming beam and the polarizing beam splitter causes phase shift Fig. \ref{fig:NLcombined}(b) and gain imbalance Fig. \ref{fig:NLcombined}(c). The incoming polarization state orientation can be controlled using a $\lambda/2$ wave plate; The wave plate orientation is permanently controlled \cite{Xu13,Zumberge04} to keep the beam aligned with the beam splitter. In Ref.~\cite{Xu13}, one wave plate adjustment technique is described. An extra beam with a known linear polarization at $\pi/4$ is injected into the interferometer. A feedback loop adjusts the angle of the wave plate to ensure that the polarization state of this reference signal is not modified by the interferometer. 
The interferometer made of the reference beam, the optical path and the polarimeter is called a polarimetric interferometer. 
In Ref.~\cite{Xu13}, the polarimetric interferometer has an accuracy of 
9~pm.

\subsection{Performance of the different interferometers}
\begin{figure}[ht!]
\centering
\begin{tikzpicture}
    \node[anchor=south west,inner sep=0] at (0,0) {\includegraphics[trim={1.2cm 0.5cm 1.2cm 0.5cm},width=\linewidth]{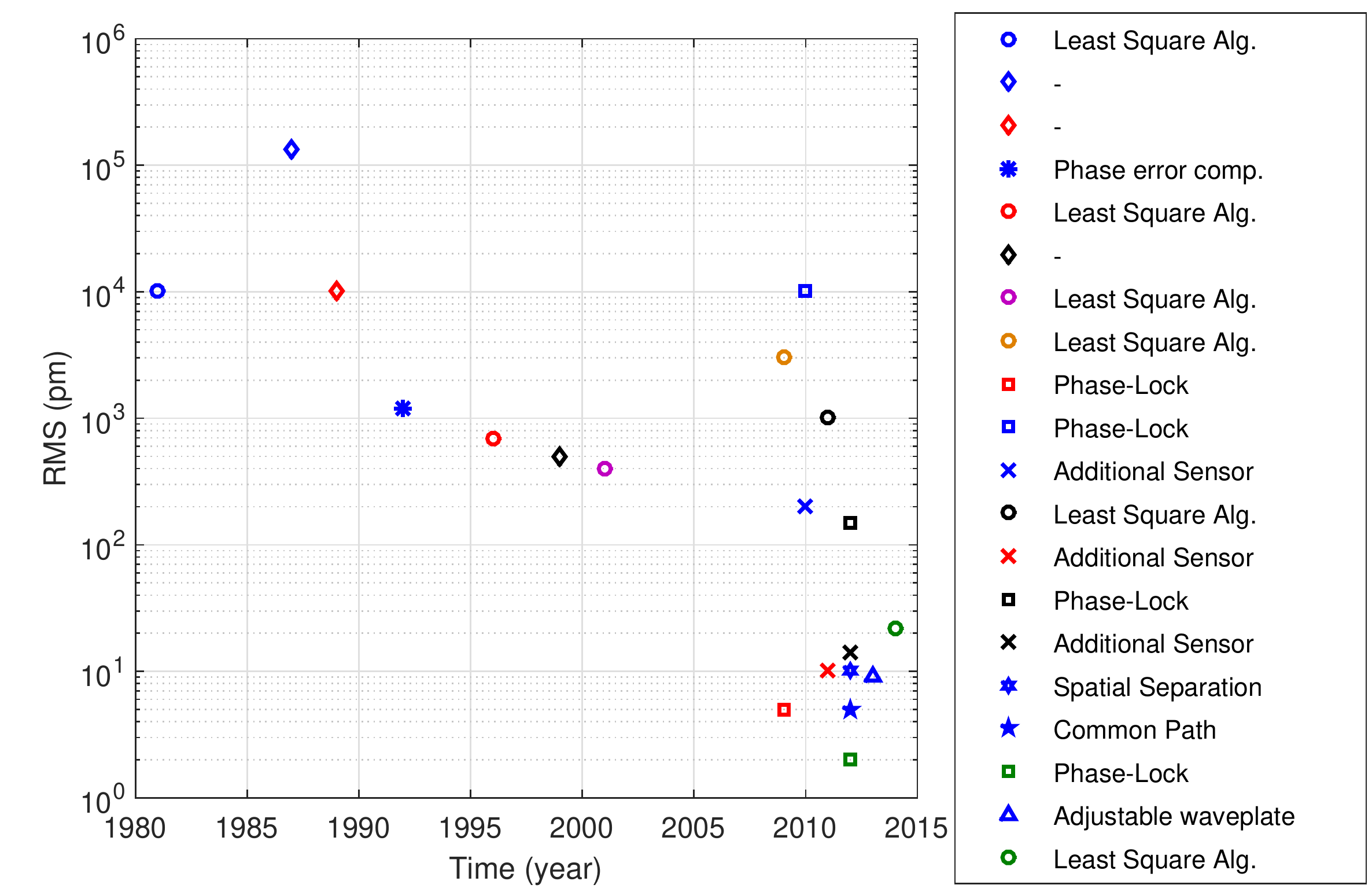}};
    \node[align=left,execute at begin node=\setlength{\baselineskip}{0.6em}] at (8.85,2.85) {\cite{Heydemann81} \\ \cite{Bobroff87} \\ \cite{Tanaka89} \\ \cite{Hou92} \\  \cite{wu96} \\ \cite{Petru99} \\ \cite{Eom01} \\ \cite{Gohlke09}  \\ \cite{Gregorcic09} \\ \cite{Xie10} \\ \cite{Korpelainen10} \\ \cite{Pozar11} \\ \cite{Seppa11}  \\ \cite{pisani12}  \\ \cite{pisani12} \\ \cite{weichert12}  \\ \cite{pisani12} \\ \cite{Celik12}   \\ \cite{Xu13} \\ \cite{Koning14}};
\end{tikzpicture}
    \caption{Time evolution of the non-linearities in RMS. The RMS values plotted are directly taken from the papers.
    The shape of the marker corresponds to an improvement or feature of the interferometer as explained in the legend. Note that the diamond marker corresponds to simple Michelson interferometers without any additional feature.}
      \label{fig:NonLinearitiesComparison}
\end{figure}

The performance of the different versions of interferometers are listed in Table~\ref{tab:nonLinearities} and chronologically represented on Fig.~\ref{fig:NonLinearitiesComparison}. The RMS of the residual non-linearities has decreased of four order of magnitude since 1980. 
After 2010, several new non-linearity reduction techniques have emerged both for homodyne and heterodyne interferometers. 
From Fig.~\ref{fig:NonLinearitiesComparison}, the Fabry-Pérot interferometer including the phase-lock method shows better results than the Michelson interferometer version.

Several papers were agreeing that the primary origin of noise comes from the non-orthogonality of the two linear polarizations measured~\cite{Hou92,Ju15,Petru99}. The use of an adjustable $\lambda /2$ wave plate can correct this issue. From Fig.~\ref{fig:NonLinearitiesComparison}, this method leads indeed to one of the lowest residual non-linearities.

\begin{table}[h]
\centering
\caption{\label{tab:nonLinearities} Residual non-linearities. Displacement error improvement by the mean of correction algorithms and other improvements are also listed. The "Real Time" column shows if the algorithm can be applied to correct in real time the error. The RMS values plotted are directly taken from the papers.} 
\begin{ruledtabular}
\begin{tabular}{ccccc}
 & &  & Residual & Real\\ 
Year&Type  &Method&displacement &Time \\
& & & error (RMS) & \\
\hline 
1981\cite{Heydemann81} & Hom. & Least square &  $10^4$ pm & no\\
1987 \cite{Bobroff87}& Hom. & & 1.32 $10^5$ pm & - \\
1996\cite{wu96} & Hom. &Least square &  700 pm & no  \\ 
1999 \cite{Petru99} & Hom.& &  $<$500 pm & no \\
2001 \cite{Eom01} & Hom. & Least square & 400 pm & yes\\
2009 \cite{Gregorcic09} & Hom. & Least square &  3 $10^3$ pm & yes \\
2010 \cite{Xie10} & Hom. &Phase-lock  & $10^4$ pm & - \\
2010 \cite{Korpelainen10} & Hom. & Capacitive reference  &  200 pm & yes\\ 
& & sensor & &  \\
2011 \cite{Seppa11} & Hom. & Capacitive reference &  10 pm & yes\\
& & sensor + improved & & \\
& &  algorithm from \cite{Korpelainen10} & & \\
2011 \cite{Pozar11} & Hom. & Least square &   $10^3$ pm & no \\
2012 \cite{pisani12} & Hom.& Common path &  5 pm & -  \\
2012 \cite{pisani12} & Hom.& Capacitive sensor corr. &  14 pm & - \\
2014 \cite{Koning14} & Hom. & Least square &  22 pm & no\\
\hline
1989 \cite{Tanaka89} & Het. &  & $<$ $10^4$ pm & no \\
1992 \cite{Hou92} & Het. & 1st order phase &  1.2 $10^3$ pm & yes \\
&&error compensation&  & \\
2009 \cite{Gohlke09} & Het. &Phase-lock &  5 pm & -  \\
2012 \cite{pisani12, weichert12} & Het.& Spatial separation  & $<$10 pm & -  \\
2012 \cite{pisani12} & Het. &Phase-lock &  150 pm & - \\
2012 \cite{Celik12,pisani12} & FPI &Phase-lock  & 2 pm & - \\
2013 \cite{Xu13} & Het.& Adjustable $\lambda /2$  & 9 pm & - \\
\end{tabular}
\end{ruledtabular}
\end{table}

\subsection{Non-linearities measurement techniques}
The RMS of the residual non-linearities have all been measured experimentally. Consequently, this section briefly summarizes the different methods used. 
In the 1980's, spectrum analyzers were used to identify the non linearities due to phase mixing in heterodyne interferometers \cite{Tanaka89}. One of the two frequencies is blocked and the remaining beat signal is measured. In case of phase mixing, 
the amplitude measured at the beat frequency provides the phase mixing amplitude. An additional measurement is performed when the two frequencies are blocked to verify that no other frequency is injected.
\\
During the next decades, the evaluation of non-linearities has led to the development of other measurement techniques. All these techniques are based on a comparison between signals: the studied one and either a reference signal or the studied signal itself but in another polarization state. 

\subsubsection*{Comparison with a reference signal}
Measurement of interferometer non-linearities is performed by comparison with a supposed error-free X-Ray Interferometer (XRI)\cite{Yacoot09,weichert12,pisani12}. The XRI used is linear to the picometre range and can work over 10~$\mu$m. 
Non-linearities are also measured by comparison between the interferometer and an \textit{error-free} Fabry-Pérot interferometer~\cite{Yin99,zhu14} or with a second Michelson interferometer~\cite{Bobroff87,Rosembluth90}.

Note that most of the residual RMS measured recently have been identified by comparison with a reference sensor.

\subsubsection*{Comparison with the signal itself}
The simplest way to identify the non-linearity is to compare the phase of the two orthogonal polarizations exiting the interferometer~\cite{Hou92}. 

Finally, the residual non-linearity can be quantified based on the Visibility parameter~\cite{Yin99}: this parameter corresponds to the difference between the maximum and minimum values measured. The bigger the Visibility parameter is, the lower non-linearity remains in the signal.

\section{Noise sources}
\label{sec:noisesources}
There are many different noise sources present in interferometry systems \cite{paschotta2008encyclopedia}. In this section, noise from the following sources will be mainly discussed: laser source, interferometer, photodiodes and data acquisition systems. The components in the interferometer are considered as perfect ones. Meanwhile, non-linearities generated by the imperfect alignment, which also appears as noise, is treated separately in the previous Section \ref{sec:nonlinearities}. The different sources of noise will be expressed in power spectral density (\rm{PSD}).

\subsection{Laser noise}
\label{lasernoise}
Different laser sources with different working principles are discussed in Ref.~\cite{csele2011fundamentals}. Among them, solid-state lasers are desirable for interferometer systems, because of their robust and compact setup, lower laser noise and long lifetime to name a few. Intensity noise and frequency noise are the main noises generated by a laser source. However, these two noises can be rejected by properly designing the interferometer: an additional photodiode to monitor the input power and normalize the signals is a good method to reduce the intensity noise, which is detailed in section~\ref{sec:AdditionalSensors}. Moreover, a well-aligned interferometer with equivalent arms length is immune to frequency noise.

\paragraph{Intensity noise}
\label{intensitynoise}
Laser intensity noise (or amplitude noise) is a typical noise generated by a single-frequency laser source. The origin of intensity noise could be various, which has been investigated in Ref.~\cite{petermann2012laser}. In practice, the relative intensity noise (\rm{RIN}), specified by laser manufacturers, is often preferred to express the intensity noise and is calculated by\cite{packard1991lightwave}

\begin{eqnarray}\label{eq:rin2} 
\rm{RIN} & = & \frac{\Phi_{I}}{<P_{0}>^2}  \quad \rm{[dB/Hz]}
\end{eqnarray}
where $\Phi_{I}$ is the PSD of the photocurrent ($\rm{W^2/Hz}$), $<P_{0}>$ is the optical power ($\rm{W}$) averaged w.r.t measurement time. 

\paragraph{Frequency noise}
\label{Frequencynoise}
Another type of noise arising from laser sources is the frequency noise, which comes from thermal effect, mechanical vibration of components, properties of the laser oscillator, etc. \cite{koechner2013solid}. Therefore, frequency noise is inherent to the emission fluctuation of single-frequency laser sources~\cite{turner2002frequency}. 
In practice, interferometers with equivalent arms are immune to frequency noise. If the two arms are not of the same length, as discussed in Section~\ref{sec:smallRange}, the frequency noise, $\Phi_{\nu}$ in unit of $\rm{Hz^2/Hz}$, appears and can be converted to displacement, $\Phi_{d}$ in the unit of $\rm{m^2/Hz}$, by Eq. 14, which is \begin{eqnarray}\label{eq:freq2disNoise}
\Phi_{d} & = & \frac{\Phi_{\nu}}{\nu^2}L_{0}^2 \quad \rm{[m^2/Hz]} \end{eqnarray} where $\nu$ is the central laser frequency ($\rm{Hz}$) and $L_{0}$ is the static arm length difference ($\rm{m}$).

\subsection{Photodiode detection system noise}
\label{pdsnoise}
The noise of a photodiode detection system includes not only the photodiodes noise but also the noise from other components in the circuits. Therefore, this section will discuss about dark current, shot noise, thermal noise on load resistance and 1/f noise on semiconductors.

\paragraph{Dark current}
\label{sec:darkCurrent}
The amplifiers used for photodectectors are of two types: photoconductive and photovoltaic~\cite{Ellis14}. Photoconductive amplifiers, like pn-junction, need a bias voltage to create the depletion region, the detection area. Consequently, the bias voltage is responsible of some current leakage called dark current (because it exists even when no light is detected). The dark current, $I_D$ can be defined as~\cite{webster1991sensors}:

\begin{eqnarray}
I_D &=& I_{SAT}(e^{\frac{qV}{k_BT}}-1) 
\end{eqnarray}
where $I_{SAT}$ is the reverse saturation current (A), $V$ is the bias voltage applied (V), $q$ is the electron charge (C), $k_B$ is the Boltzmann's constant (J/K) and $T$ is the temperature (K).

Photovoltaic detectors, on the other hand, do not require a bias voltage. Consequently, they do not present any current leakage.

\paragraph{Shot noise}
\label{shotnoise}

Shot noise, or Schottky noise, is caused by the discrete nature of photons and electric charges across potential barriers, such as diodes, transistors or p–n junctions~\cite{webster1991sensors}. Two photons with the same energy will not create the same number of electron-hole pairs. Consequently, the photocurrent generated is fluctuating. The PSD of the shot noise is given by
\begin{eqnarray}\label{eq:shotNoise} 
\Phi_{S} & = & 2qI_{\rm{PD}} \quad \rm{[A^2/Hz]}
\end{eqnarray}
where $I_{\rm{PD}}$ is the average photocurrent ($\rm{A}$) that crosses the barrier, $q$ is the electron charge ($\rm{C}$). From the equation, we can say that shot noise is a white noise. Moreover, as it is proportional to the photocurrent value, a higher current causes more random motion which leads to higher shot noise.

\paragraph{Thermoelectrical noise}
\label{Thermoelectricalnoise}
Thermoelectrical noise or Johnson noise \cite{johnson1928thermal} is generated by the thermal fluctuation of electrons passing through resistive components of the sensor circuits. The PSD of thermoelectrical noise is given by
\begin{eqnarray}\label{eq:johnson} 
\Phi_{T} & = & 4k_{\rm{B}}TZ_{\rm{R}} \quad \rm{[V^2/Hz]}
\end{eqnarray}
where $k_{\rm{B}}$ is the Boltzmann's constant ($\rm{J/K}$), $T$ is the Kelvin temperature ($\rm{K}$) and $Z_{\rm{R}}$ is the equivalent resistance ($\rm{\Omega}$) of the whole system. The equation shows that thermoelectrical noise is a white noise. Also, it depends on the temperature and the resistive load of the circuit.

\paragraph{1/f noise}
\label{1/fnoise}
Flicker noise, or 1/f noise, corresponds to fluctuation in the resistance of semiconductors and occurs in all electronic components~\cite{keshner19821,webster1991sensors}. The main characteristic of the 1/f noise is that its power spectral density is inversely proportional to the frequency. The model of the 1/f noise can be expressed as
\begin{eqnarray}\label{eq:Flicker} 
\Phi_{1/f} & = & K/f^a \quad \rm{[V^2/Hz]}
\end{eqnarray}
where $K$ is a constant related to the circuit, $f$ is the frequency and $a$ is a coefficient between 0 and 2, and usually close to 1.

\subsection{Data acquisition system noise}
\label{quantizationnoise}
Data acquisition system (\rm{DAQ}) and its Analog to Digital Converters (\rm{ADC}) have a certain noise floor, which is related to the input referred noises and its quantization noise. The sources of input referred noise in the data acquisition system are similar to the sources discussed in the previous section~ \ref{pdsnoise}. If the sampling frequency and the bits of the ADC is not high enough, its quantization noise dominates the noise floor. The quantization noise induced by the ADC can be measured by disconnecting all other inputs and outputs from the ADC and recording the signal directly. The PSD of the theoretical ADC noise is given by 
\begin{eqnarray}\label{eq:adcNoise} 
\Phi_{ADC} & = & \frac{q^2}{{12f_{n}}} \quad \rm{[V^2/Hz]}
\end{eqnarray}
where $q=2\Delta V/2^{n+1}$ is the quantization interval, $\Delta V$ is the half of voltage range, $n$ is the number of bits available to the Data Acquisition card (DAQ) and $f_{n}$ is the Nyquist frequency~\cite{oppenheim1999discrete}. 

In addition, the sampling time and the time processing to extract the phase induce some delay~\cite{Ellis14}. This delay induces an uncertainty on the phase measured and consequently on the displacement of the moving mirror at a certain time. The error on the displacement is called the data age error and is larger if the speed of the mirror is higher~\cite{Ellis14}.

\subsection{Ambient noise}
\label{Ambientnoise}
Fluctuations of temperature and pressure are responsible of signal variations and can thus be considered as an additional source of noise. In fact, they modify the refractivity of the air which makes the optical path lenght vary~\cite{Bobroff93}. In order to reduce the temperature influence, the interferometer can be placed inside a vacuum chamber. Another option is to use a weather station and correct the signal based on the pressure, temperature and humidity measurements~\cite{Kren09,pisani12,Xu13}. In addition, the optical elements have to be placed in a compact~\cite{Wu2002} monolithic block made of a material with a low thermal expansion coefficient e.g. Zerodur, fused silica \cite{Schuldt07}. The resolution reached with this last improvement is lower than 5 pm/$\sqrt{\rm Hz}$ above 10~mHz~\cite{Schuldt12}. 
\\
Ambient light is also responsible of some spurious current injected in the photodiodes. As we are discussing about compact devices, it will be easy to reject this ambient light by putting the interferometer in the dark.
\\
Finally, electronics are responsible of acoustic noise. To avoid its influence, the electronics have been placed in another room in Ref.~\cite{Yacoot09,pisani12}. 
\subsection{Overall noise model}
\label{allnoise}
\begin{figure}[ht!]
    \centering
	\includegraphics[trim = {0cm 0cm 0cm 0cm},width=1\linewidth]{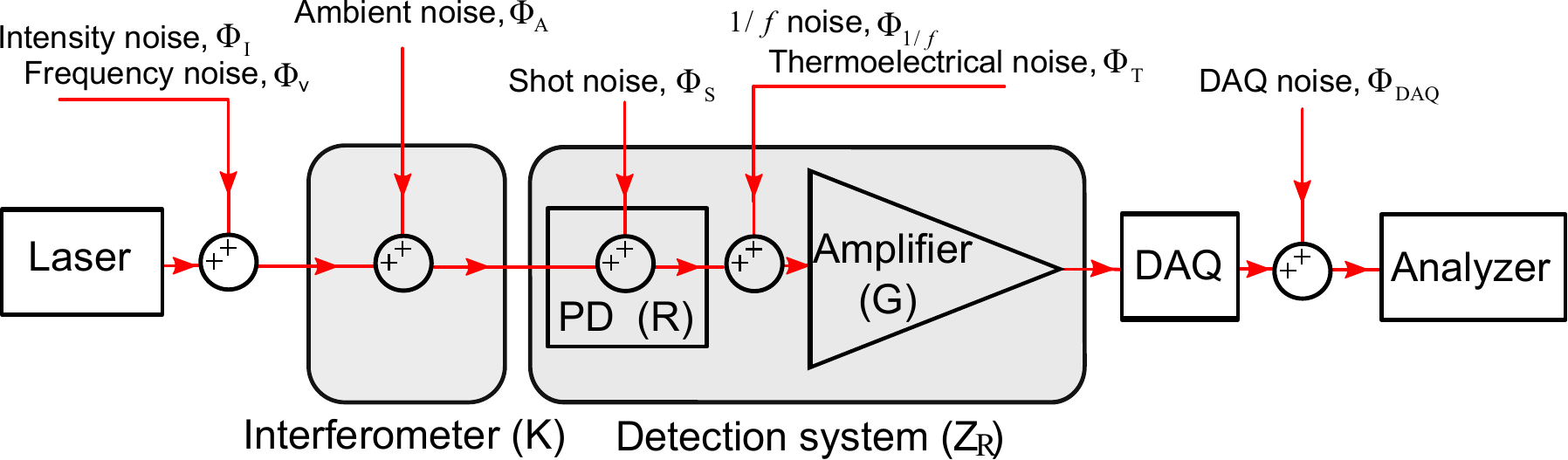}
	\caption{Noises added to the interferometer system. The red arrow is the flow of the laser, current or data. $R$ is the responsivity of the photodiodes ($\rm{A/W}$), which converts the laser power into current. $Z_{\rm{R}}$ is the equivalent impedance of the circuit ($\rm{\Omega}$) and $G$ is the load resistance of the amplifier ($\rm{\Omega}$), which is also the gain of the circuit.}
	 \label{fig:allnoise}
\end{figure}
The flowchart of the noise model including the sources of noise mentioned is shown in Fig.~\ref{fig:allnoise}. The model is based on several assumptions. The first one is that the sources of electronic noise are uncorrelated. The second one is that the input of the amplifier is the photocurrent and the output signal is the voltage. The third one is that the ambient noise $\Phi_{A}$ is simplified as optical power fluctuation in the interferometer. Moreover, the filters of the circuits are excluded.
From the left side to the right side, a laser beam containing intensity noise and frequency noise is generated by the laser source and then enters the interferometer. The ambient noise is added inside the interferometer and can be seen as a laser power fluctuation. On the photodiode, they are converted into a current fluctuation by $R$. Moreover, the shot noise, which is a current, is generated in the photodiode. before being amplified, the currents corresponding to the thermoelectrical noise and 1/f noise are added. The current fluctuation is converted into voltage fluctuation by the gain of the amplifier $G$. When the data is recorded by the DAQ, the DAQ noise is added to the noise floor as well. The overall noise $\Phi_{total}$ in a consistent unit can be expressed as
\begin{eqnarray}\label{eq:Noisemodel}
\Phi_{total} &=& G^2\{R^2[\Phi_{I}+\Phi_{A}]+\Phi_{S}
\nonumber\\
&+&Z_{\rm{R}}^{-2}(\Phi_{1/f}+\Phi_{T})\} +\Phi_{DAQ} \quad \rm{[V^2/Hz]} 
\end{eqnarray} 

In the analyzer, the noise unit is converted from $\rm{V^2/Hz}$ to $\rm{m^2/Hz}$ by the data processing methods. The different data processing methods corresponding to the different types of interferometers are introduced in Section~\ref{sec:smallRange}, Section~\ref{sec:homodyne} and Section~\ref{sec:heterodyne}.

\section{Summary}
\label{sec:discussion}

\paragraph*{} This paper has presented a review of `compact' interferometers that employ different methods to increase the dynamic range compared with that of a simple interferometer. All techniques are based on the same principle: create a phasemeter by generating two (or more) quadrature signals from which the phase, and as such the displacement, can be extracted over more than one fringe by unwrapping the outputs with a 4-quadrant arctangent. 

To determine the size of systems, we searched for their dimensions in the literature. From Table~\ref{tab:resolutionEvolutionHomodne}, we see that in average the optical homodyne interferometer occupies an area of approximately 17x17~cm$^2$, with some substantial variation in size. Heterodyne phasemeters are somewhat larger, typically 30x30\,cm$^2$, but in both cases the `size' often neglects the input beam preparation optics and data acquisition system. Heterodyne devices typically require more space as either an additional laser source or an AOM is required.

In most homodyne systems, two polarization states are used to sample the target mirror with different phase shifts, creating the quadrature outputs. For heterodyne interferometers, two beams with different laser frequencies pass through the interferometer and the quadrature signals are most commonly created by demodulating the beat signal with a sine and a cosine.

\paragraph*{}The resolution of homodyne phasemeters has improved considerably since their inception, largely due to decreasing technical noises. Table~\ref{tab:resolutionEvolutionHomodne} shows that several devices in the last 10 years have reached a sensitivity at or below 1\,pm$/\sqrt{\rm{Hz}}$ at 1\,Hz. To improve sensitivity it is possible to increase the number of reflections in one or both arms of an interferometer (Section~\ref{sec:MultipassHomodyne}). Several experiments have employed additional photodiodes to reduce intensity-noise coupling (Section~\ref{sec:HomodyneExtraphotodiodes}).

\paragraph*{}Heterodyne phasemeters push (much of) the optical complexity onto the input beam preparation and the signal analysis. It is difficult to make a fair comparison between devices to the very large range of design parameters, including the heterodyne frequency and the measurement frequency, but Table~\ref{tab:hetSummary} shows that resolutions less than 1\,pm$/\sqrt{\rm{Hz}}$ are routinely achieved. Many of the devices reviewed were not very compact, and it is difficult to determine the size scale of the complete apparatus, including the input-beam preparation. 

\paragraph*{} Overall, heterodyne interferometers are larger and more complex than their homodyne cousins, but they are less susceptible to low-frequency readout noise. A brief summary of the common noise sources that limit the resolution of interferometers is included in Section~\ref{sec:noisesources}.

\paragraph*{} A significant advantage of all phasemeters is that they are inherently calibrated to the wavelength of the laser. There are, however, several sources of non-linearity that affect their accuracy, and these have also been reviewed. Non-linearities can be reduced in several ways. Ellipse fitting algorithms, section~\ref{sec:EllipseFittingAlgorithm}, are widely used to transform the phasemeters output into a unitary circle centred at the origin, removing both the leading order of non-linearity and the offsets inherent to measuring intensity with photodiodes. Additional sensors, section~\ref{sec:AdditionalSensors}, can reduce the residual non-linearity by reducing the effect of power fluctuations or by subtracting large input phase-shifts. Polarization mixing (homodyne) and phase mixing (heterodyne) can be reduced thanks to spatially separating beams, section~\ref{sec:PhaseMixing}. The most promising heterodyne technique for reducing non-linearity is to employ a `phase lock' to hold the signal in a single quadrature and read out the phase-shift required to keep it there. From figure~\ref{fig:NonLinearitiesComparison}, it is clear that non-linearities have been improving consistently during the last decades and that modern interferometers can consistently achieve single-digit picometer accuracy.

\paragraph*{} Heterodyne interferometers achieve a good resolution and can inherently reject any DC component. However, there is still room for improvement before they reach the level of compactness of homodyne devices. As explained in this Section, the size of the beam preparation equipment limits the compactness. Therefore, reducing the size of these devices should be the concern of future research if they want to compete with the size of the homodyne devices.
\paragraph*{} In addition, many designs have already been developed that offer significant improvement in resolution and sensitivity. Nevertheless, whether for homodyne or heterodyne interferometers, a lot of parameters change from one device to another (wavelength, type of photodetectors, etc.). In order to have a fair comparison between the different solutions proposed, they should be tested on the same setup. Such a prototype should also allow to see if certain solutions can be combined and whether the performance of the combined interferometer corresponds to the combination of the performance obtained with the two solutions independently.

\begin{acknowledgments}
This project has received funding from the European Union’s Horizon 2020 research and innovation programme under the Marie Sklodowska-Curie grant agreement No 701264. In addition, F.R.S.-FNRS is acknowledged for supporting this research under the incentive grant (grant agreement No F.4536.17). The authors gratefully acknowledge the CSC for funding Binlei Ding, grant agreement No 201607650017.
The authors gratefully acknowledge the members of the LSC seismic working group, and in particular Dr. Fabrice Matichard, for the fruitful collaboration and inspiring discussions. 
This paper has been assigned the LIGO document number LIGO-P1700102.
\end{acknowledgments}


%

\end{document}